\documentclass[11pt,a4paper]{article}
\PassOptionsToPackage{hypertexnames=false}{hyperref}
\pdfoutput=1
\usepackage{jheppub}
\makeatletter
\gdef\@fpheader{}
\makeatother
\usepackage{bm}
\usepackage{mathtools}
\usepackage{booktabs}
\usepackage{array}
\usepackage{tabularx}
\usepackage{graphicx}
\usepackage{subcaption}
\usepackage{mathrsfs}
\usepackage{slashed}
\usepackage{microtype}
\usepackage{float}
\usepackage{tikz}
\usetikzlibrary{arrows.meta,calc,positioning,decorations.pathmorphing}
\usepackage{comment}

\newcommand{\dd}{\mathrm{d}}
\newcommand{\ee}{\mathrm{e}}
\newcommand{\ii}{\mathrm{i}}
\newcommand{\Tr}{\operatorname{Tr}}
\newcommand{\SK}{\mathrm{SK}}
\newcommand{\IF}{\mathrm{IF}}
\newcommand{\DR}{\mathcal{D}_{R}}
\newcommand{\NN}{\mathcal{N}}
\newcommand{\EE}{\mathcal{E}}
\newcommand{\BB}{\mathcal{B}}

\newcommand{\avg}[1]{\left\langle #1\right\rangle}

\newcommand{\ord}{\mathcal{O}}

\newcommand{\beq}{\begin{equation}}
\newcommand{\eeq}{\end{equation}}

\title{Schwinger--Keldysh worldline effective field theory for quantum black holes:~dissipation, noise and decoherence}

\author[a]{Hajime Kobayashi}
\affiliation[a]{Center for Gravitational Physics and Quantum Information, Yukawa Institute for Theoretical Physics, Kyoto University, 606-8502, Kyoto, Japan}
\emailAdd{hajime.kobayashi@yukawa.kyoto-u.ac.jp}

\preprint{YITP-26-147}

\abstract{
We formulate a Schwinger--Keldysh worldline effective field theory for quantum black holes that treats dissipation, fluctuations, and decoherence within one real-time influence functional. The reduced dynamics is encoded in causal response and symmetric noise kernels. Their spectra obey positivity constraints, while thermal Kubo--Martin--Schwinger symmetry relates them through the fluctuation--dissipation relation. A Hubbard--Stratonovich representation turns the noise kernel into a Gaussian source driving the resolved metric and worldline. For a leading post-Newtonian Schwarzschild binary, direct and metric-mediated stochastic forces combine into the complete relative-acceleration noise at this order, preserve center-of-mass momentum balance, and remain Planck suppressed. The same noise kernel gives the quadrupolar decoherence functional for a supported spatial superposition and reproduces the Danielson--Satishchandran--Wald horizon-induced scaling after absorptive matching. Away from thermal equilibrium, response and noise remain independent matching data. We discuss extensions to microscopic quantum structure in the worldline kernels, matching to spatially extended coarse-grained descriptions, and gravitational splitting.
}

\begin{document}
\maketitle
\flushbottom

\section{Introduction}
\label{sec:intro}

Black holes carry Bekenstein--Hawking entropy and emit Hawking radiation, suggesting that the classical exterior geometry is a coarse-grained description of underlying quantum degrees of freedom~\cite{Bekenstein:1973ur,Hawking:1975vcx}.\footnote{For complementary on-shell approaches to black-hole thermality, see e.g. Refs.~\cite{Aoude:2024hawking,Aoki:2025quantum,Akpinar:2026ldb}.  Here the black-hole correlators are instead treated as input to an exterior reduced theory; see also the discussion around Eq.~\eqref{eq:detailed_balance}.}  Because the horizon hides the interior, an exterior observer cannot directly resolve the detailed state of these horizon/internal degrees of freedom.  The question is then how much information about this unresolved sector can be inferred indirectly from its response to external tidal fields and accompanying fluctuations.

At wavelengths much longer than the black-hole size, this question can be formulated within worldline effective field theory (EFT), in which the unresolved black hole is represented by localized multipole operators coupled to exterior tidal fields~\cite{Goldberger:2004jt,Goldberger:2005cd,Porto:2016pyg}.  Their correlation functions characterize how the unresolved sector responds to, and fluctuates under, long-wavelength exterior probes.  In binaries, the absorptive part of this response corresponds to horizon absorption of the companion-induced perturbation, commonly called \emph{tidal heating} or \emph{tidal dissipation}~\cite{Poisson:2004absorption,Chia:2020yla,Saketh:2022xjb}.\footnote{For reviews of finite-size and dissipative effects in compact-object dynamics, see Refs.~\cite{Porto:2016pyg,Levi:2020eft,CardosoPani:2019review,Sherf:2021tidal}.  Representative treatments range from classical horizon-flux calculations~\cite{PoissonSasaki:1995,Alvi:2001,Chatziioannou:2012heating} to absorptive worldline EFT and modern response/scattering formulations~\cite{Goldberger:2005cd,Porto:2007absorption,CreciHindererSteinhoff:2021,IvanovZhou:2023matching,KatagiriNakanoOmukai:2023,Kobayashi:2025tdn}.  For using tidal heating to constrain departures from perfect absorption, including near-horizon or surface reflectivity models, see Ref.~\cite{Datta:2019euh}.}  In the standard worldline EFT, retarded multipole correlators are fixed by low-frequency absorption and by nonzero-frequency matching of the metric one-point function to black-hole perturbations~\cite{Goldberger:2005cd,IvanovZhou:2023matching,Kobayashi:2025tdn,CombaluzierSzteinsznaider:2025dyn,Kobayashi:2025vgl}.

Absorptive response, however, is only one part of the reduced quantum dynamics.  The open-system split adopted here treats the horizon/internal black-hole degrees of freedom as the unresolved sector, while the exterior tidal and probe sectors remain explicit.\footnote{A distinct class of worldline-EFT studies instead considers dissipative effects associated with an \emph{external} environment of a compact object~\cite{Modrekiladze:2024fluid,Modrekiladze:2026selfforce,Modrekiladze:2026media,Bern:2026environment,WilsonGerow:2026friction}.}  Tracing over the unresolved sector can then generate fluctuations, stochastic backreaction and loss of coherence in the reduced exterior state~\cite{Feynman:1963fq,Calzetta:1986ctp,Hu:2008rga}.  In thermal equilibrium these fluctuations are constrained by the Kubo--Martin--Schwinger (KMS) condition and the associated fluctuation--dissipation theorem (FDT), but this relation is state dependent; away from equilibrium the fluctuation data need not be determined by the causal response~\cite{Kubo:1966nj,Kamenev:2011,Sieberer:2015hba}.  This motivates a real-time extension of worldline EFT that places dissipation and fluctuations on the same footing while preserving the standard low-energy matching logic.

The natural framework for this extension is the closed-time-path (CTP), or Schwinger--Keldysh (SK), effective action~\cite{Schwinger:1960qe,Keldysh:1964ud,Feynman:1963fq}.  We refer to the resulting construction as the \emph{SK worldline EFT}.
This should also be distinguished from bottom-up ``open EFT'' programs that construct local SK actions from microscopic unitarity, locality, and the doubled symmetry structure of the contour~\cite{Crossley:2015evo,Glorioso:2017fpd,Hongo:2019open,Akyuz:2024coset,Lau:2025gravity}.\footnote{The symmetry-based program has been extended to gauge and gravitational systems, where the interplay between physical and ``advanced'' SK symmetries constrains dissipation and noise~\cite{AguiSalcedo:2025light,Christodoulidis:2026consistency,AbeNishii:2026nonabelian}.  In cosmology it underlies open EFTs of inflation and gravity, with applications to primordial non-Gaussianity, cosmological-collider signals, cosmic microwave background (CMB) constraints on dissipative inflation, gravitational-wave propagation and dark-energy phenomenology~\cite{AguiSalcedo:2024inflation,Colas:2025cosmocollider,AguiSalcedo:2026dissipative,AguiSalcedo:2025gravity,AguiSalcedo:2026darkenergy}.}  For electric- and magnetic-type tidal sources $\mathcal X^{\EE,L}\equiv\EE^L$ and $\mathcal X^{\BB,L}\equiv\BB^L$, integrating out the unresolved black-hole variables gives, at quadratic order,
\begin{equation}
  S_{\IF}^{(2)}=
  \sum_{A,B,L,L'}\int\dd\tau\dd\tau'\,
  \mathcal X^{A,L}_{a}(\tau)
  \left[
    \DR^{AB}{}_{LL'}(\tau,\tau')\mathcal X^{B,L'}_{r}(\tau')
    +\frac{\ii}{2}\NN^{AB}{}_{LL'}(\tau,\tau')\mathcal X^{B,L'}_{a}(\tau')
  \right].
  \label{eq:intro_IF_summary}
\end{equation}
Here and below the sums run over $A,B\in\{\EE,\BB\}$ and $\ell,\ell'\geq2$, with symmetric trace-free (STF) indices implicit in $L,L'$.  The conventional multipolar combinatorial factors are absorbed into the definition of $Q_A^L$, so the quadrupole is unit-coupled.  The retarded kernel $\DR$ controls response and dissipation, whereas the symmetric kernel $\NN$ controls fluctuations, stochastic backreaction and decoherence; they remain independent until a state condition such as KMS is imposed.

The CTP construction separates the coarse-graining procedure from assumptions about the initial state of the unresolved black-hole sector.  The response and noise kernels are evaluated in the source-free state $\rho_{\rm BH}^{\rm init}$ and then probed by the exterior tidal fields.  For a Gibbs choice of $\rho_{\rm BH}^{\rm init}$ under a stationary black-hole Hamiltonian, the correlators satisfy the KMS condition and FDT; the subsequently tidally perturbed state need not remain thermal~\cite{Kubo:1966nj}.  With the appropriate antiunitary reversibility operation, the same equilibrium structure appears in a local SK action as dynamical KMS symmetry~\cite{Crossley:2015evo,Glorioso:2017fpd,Akyuz:2024coset,Lau:2025gravity}.  For Schwarzschild equilibrium the temperature is fixed to the Hawking temperature.  Keeping response and noise independent, and retaining two-time kernels when needed, also accommodates nonthermal or nonstationary choices of $\rho_{\rm BH}^{\rm init}$.

The formalism has two immediate applications.  First, the symmetric kernel admits the standard Hubbard--Stratonovich (HS) representation, in which the imaginary quadratic influence action is written as an average over an auxiliary Gaussian field $\xi_A^L$ with covariance $\NN$~\cite{Stratonovich:1957,Hubbard:1959,Feynman:1963fq,Hu:2008rga}.  Such stochastic representations are standard in influence-functional and stochastic-gravity treatments~\cite{Sinha:2002stoch,Hu:2008rga}, and stochastic worldline equations are known for particles coupled to quantum fields~\cite{GalleyHu:2005}; dissipative black-hole worldline EFT, meanwhile, matches intrinsic multipole response to horizon absorption~\cite{Goldberger:2005cd}.  Our new step is to combine these ingredients while keeping both the exterior metric and black-hole worldline resolved.  The same $\xi_A^L$ therefore drives a coupled metric--worldline system, generating metric-mediated stochastic forces and metric--orbit cross correlations.  At leading post-Newtonian order this also gives a complete relative-acceleration noise source for a nonspinning binary, including the reaction required for center-of-mass momentum balance.

Second, the Danielson--Satishchandran--Wald (DSW) proposal gives a complementary reason to study the fluctuation sector: decoherence is controlled by the symmetric correlator rather than by causal response alone.  A probe held in a spatial superposition outside a black hole can imprint branch-dependent information on horizon/soft gravitational degrees of freedom, so tracing over that record suppresses the branch coherence~\cite{Danielson:2022tdw,Danielson:2023yhu,Danielson:2024yru,Wilson-Gerow:2024ljx}.  Thermal-correlator and holographic descriptions support the same infrared-sensitive picture~\cite{Biggs:2024doh,Kawamoto:2025decoherence}.  Our contribution is to embed this mechanism in the same matched worldline EFT used for response and stochastic backreaction: the DSW exponent is the noise kernel $\NN$ contracted with the difference between two exterior tidal histories.  Equilibrium FDT then relates decoherence to absorption, while away from equilibrium $\NN$ remains independent matching data.  DSW decoherence and noisy orbital dynamics thus probe complementary aspects of one low-energy black-hole fluctuation spectrum.

Beyond these explicit applications, the SK worldline EFT points to three broader directions in which its matching data can probe black-hole quantum dynamics.  First, microscopic near-horizon models, including Jackiw--Teitelboim (JT)/Schwarzian and finite-entropy descriptions, can determine how quantum structure enters two-point kernels, higher correlators and nonequilibrium evolution~\cite{Maldacena:2016upp,Iliesiu:2020qvm,Saad:2019lba,Stanford:2019vob}.  Quantum-chaotic descriptions provide a complementary diagnostic: the eigenstate thermalization hypothesis (ETH) organizes the smooth, thermal-like matrix-element data visible in two-point kernels, whereas level correlations, out-of-time-order correlators (OTOCs) and higher-point observables probe finer microstate information~\cite{Deutsch:1991,Srednicki:1994,Deutsch:2018eth,Maldacena:2015waa}.  Second, the worldline EFT can be matched to spatially extended coarse-grained descriptions in which the effective degrees of freedom and their interactions retain extended spatial support.  Stretched-horizon and finite-cutoff fluid descriptions retain spatial transport and gradients along a hypersurface~\cite{Price:1986yy,Thorne:1986membrane,Gourgoulhon:2005dns,Bredberg:2011wilsonian}, whereas the Einstein--Langevin framework retains stochastic correlations and propagation in the bulk curved spacetime~\cite{Calzetta:1986ctp,Hu:2008rga,Sinha:2002stoch,HuRoura:2007}.  The worldline EFT still retains finite-size effects through multipoles and Wilson coefficients; the additional matching asks how this extended spatial structure reduces to localized worldline data.  Third, any reduced influence-functional description in gravity requires a gravitational splitting: a specification of the observables retained and of the gravitational degrees of freedom traced over~\cite{Donnelly:2016auv,Speranza:2017gxd}.  This becomes especially relevant for decoherence, where the subsystem choice determines which environmental records are excluded from the reduced description.  These broader directions are discussed further in Section~\ref{sec:outlook}.

The paper is organized as follows.  Section~\ref{sec:worldline} derives the SK worldline EFT, its positivity properties, the connection to ETH, and the Hubbard--Stratonovich representation.  Section~\ref{sec:kms} develops equilibrium KMS/FDT and dynamical KMS, then distinguishes stationary nonthermal states from genuinely nonstationary two-time kernels.  Section~\ref{sec:applications} discusses noise-driven backreaction on the resolved metric and worldline, and DSW decoherence.  Section~\ref{sec:outlook} considers microscopic quantum structure in the worldline kernels, matching to spatially extended coarse-grained descriptions, and gravitational splitting and subsystem definition.  Section~\ref{sec:conclusions} presents concluding remarks.

\paragraph{Conventions and notation.}
Unless otherwise stated we use $G=c=\hbar=k_{\rm B}=1$ and signature $(-,+,+,+)$; factors of $\hbar$, $c$ and $k_{\rm B}$ are restored when needed.  Greek indices are spacetime indices, $i,j,\ldots$ label a local orthonormal spatial frame, and $L=i_1\cdots i_\ell$ denotes an STF multi-index.  $A,B\in\{\EE,\BB\}$ label electric- and magnetic-type tidal channels.  For operators, $[X,Y]\equiv XY-YX$ and $\{X,Y\}\equiv XY+YX$ denote the commutator and anticommutator, respectively.  We use $r_h$ for the characteristic horizon radius (for Schwarzschild, $r_h=2GM$).  CTP branches are labeled $1,2$ and average/difference variables by $r,a$.  Worldline kernels use proper time $\tau$; $t$ (and the protocol duration $t_f$) denotes exterior time. Their leading weak-field identification is specified in Section~\ref{subsec:decoherence_general}.  Our Fourier convention is
\begin{equation}
  F(s)=\int_{-\infty}^{\infty}\frac{\dd\omega}{2\pi}\,
  e^{-\ii\omega s}F(\omega),
  \qquad
  F(\omega)=\int_{-\infty}^{\infty}\dd s\,
  e^{+\ii\omega s}F(s).
  \label{eq:fourier_convention_intro}
\end{equation}

\section{Worldline EFT on the closed time path}
\label{sec:worldline}
\label{sec:sk}

\subsection{Degrees of freedom and regime of validity}

Let $g_{\mu\nu}$ denote the long-wavelength metric resolved by the EFT and let a compact object of characteristic size $r_\star$ follow a worldline $z^\mu(\tau)$.  Replacing the short-distance structure inside a worldtube of radius $\sim r_\star$ by worldline degrees of freedom is justified, for a black hole with $r_\star\sim r_h$, when
\begin{equation}
  r_h/\mathcal{R}\ll 1,
  \qquad
  r_h\omega\ll 1,
  \label{eq:powercounting}
\end{equation}
where $\mathcal R$ is the characteristic curvature, orbital-separation or external-field scale and $\omega$ a resolved frequency~\cite{Goldberger:2004jt,Goldberger:2005cd,Porto:2016pyg}.  For perturbative calculations one may further split
\begin{equation}
  g_{\mu\nu}=\bar g_{\mu\nu}+\frac{1}{M_{\rm Pl}}h_{\mu\nu},
  \label{eq:background_split}
\end{equation}
For the post-Newtonian (PN) inspiral applications emphasized in this paper we take the reference background to be Minkowski, $\bar g_{\mu\nu}=\eta_{\mu\nu}$, and organize the long-wavelength gravitational field in the usual weak-field expansion around it.  The worldline coefficients then encode the compact-object structure that has been integrated out, while the binary potential and radiation fields are retained explicitly.  This specialization is a choice of calculational regime rather than a limitation of the point-particle EFT itself.\footnote{The covariant worldline construction can equally be formulated for a compact object moving in an arbitrary curved background.  In particular, a small compact object---including a small black hole---in an extreme-mass-ratio inspiral (EMRI) can be treated as an effective point particle on the strong-field geometry of the large black hole without invoking a weak-field or slow-motion expansion of that background~\cite{GalleyHu:2009}.}

For a quasi-circular compact-binary inspiral one has $\omega\sim m\Omega$ for an orbital harmonic and $\mathcal R\sim r_{12}$ in the PN regime.  Each harmonic samples the compact-object response at the corresponding orbital frequency: analytic real terms generate conservative tidal corrections, the absorptive part generates horizon flux/tidal heating (tidal dissipation), and nonzero-frequency one-point-function matching encodes dynamical tidal response~\cite{Goldberger:2005cd,Goldberger:2020yvj,Chia:2020yla,Saketh:2022xjb,Chia:2024nob,CombaluzierSzteinsznaider:2025dyn,Kobayashi:2025vgl}.  The SK worldline EFT below supplements this deterministic retarded sector by the fluctuation data of the same coarse-grained black-hole degrees of freedom.

Before integrating out that sector, the generic point-particle action contains the mass and rotational degrees of freedom,
\begin{equation}
  S_{\rm pp}[z,g,S]
  =-M\int \dd\tau + S_{\rm spin}[z,g,S],
  \label{eq:Spp}
\end{equation}
where $S_{\rm spin}$ contains the rotational degrees of freedom and their couplings.  We henceforth specialize to the nonspinning sector, $S=0$, and write $S_{\rm pp}[z,g]\equiv S_{\rm pp}[z,g,S=0]$; spin and spin-induced finite-size operators will therefore not be presented explicitly.  Finite-size response in the sector considered here is generated by integrating out the unresolved black-hole variables rather than by introducing a separate local finite-size action.

For this nonspinning black hole, let $X$ denote the unresolved variables and $S_X[X]$ their source-free action, with Hamiltonian $H_{\rm BH}$.  We define the source-free operator evolution by $Q_A^L(\tau)=e^{\ii H_{\rm BH}(\tau-\tau_i)}Q_A^L e^{-\ii H_{\rm BH}(\tau-\tau_i)}$, where $A\in\{\EE,\BB\}$ and $L=i_1\cdots i_\ell$.  We take the reference black-hole sector to be spherically symmetric, with rotational- and parity-invariant dynamics.  For the rotationally invariant initial states considered below,
\begin{equation}
  \avg{Q_A^L(\tau)}_{\rho_{\rm BH}^{\rm init}}=0,
  \qquad A\in\{\EE,\BB\},\quad \ell\geq2,
  \label{eq:zero_mean}
\end{equation}
where $\avg{\mathcal O(\tau)}_{\rho_{\rm BH}^{\rm init}}\equiv\Tr_{\rm BH}[\rho_{\rm BH}^{\rm init}\mathcal O(\tau)]$ for any source-free black-hole operator.  Thus Eq.~\eqref{eq:zero_mean} states that the reference one-point functions vanish, while the corresponding two-point functions can remain nonzero.

The interaction with the long-wavelength gravitational field is organized as a multipolar expansion in tidal tensors.  At leading order in the tidal fields, rotations fix the STF contractions and parity forbids electric--magnetic cross couplings, giving~\cite{Goldberger:2005cd}
\begin{equation}
  S_{\rm int}[z,g,X]
  =\sum_{\ell\geq 2}\int \dd\tau
  \left[
    Q_{\EE}^L[X](\tau)\,\EE_L(\tau)
    +Q_{\BB}^L[X](\tau)\,\BB_L(\tau)
  \right].
  \label{eq:tidal_coupling}
\end{equation}
At quadrupolar order, $\EE_{ij}$ and $\BB_{ij}$ are the electric and magnetic parts of the Weyl tensor $C_{\mu\alpha\nu\beta}[g]$ computed from the long-wavelength perturbed metric $g_{\mu\nu}$ in~\eqref{eq:background_split},
\begin{equation}
  \EE_{ij}=C_{\mu\alpha\nu\beta}
   e_i^{\mu}u^{\alpha}e_j^{\nu}u^{\beta},
  \qquad
  \BB_{ij}=\frac12\epsilon_{i}{}^{kl}
  C_{\mu\alpha\nu\beta}e_k^{\mu}e_l^{\alpha}e_j^{\nu}u^{\beta},
  \label{eq:tidal_tensors}
\end{equation}
with $u^\mu=\dd z^\mu/\dd\tau$ and an orthonormal spatial triad $e_i^\mu$.  These tensors pull the ambient Weyl field back to the worldline; varying $z$ probes its neighborhood.  Higher-rank moments are the STF-projected spatial derivatives
\begin{equation}
  \EE_{i_1\cdots i_\ell}
  =D_{\langle i_3}\cdots D_{i_\ell}\EE_{i_1 i_2\rangle},
  \qquad
  \BB_{i_1\cdots i_\ell}
  =D_{\langle i_3}\cdots D_{i_\ell}\BB_{i_1 i_2\rangle},
  \qquad \ell\geq3,
  \label{eq:higher_tidal_tensors}
\end{equation}
where $D_i\equiv e_i{}^\mu P_\mu{}^\nu\nabla_\nu$, $P_\mu{}^\nu=\delta_\mu{}^\nu+u_\mu u^\nu$, and brackets denote STF projection.  We absorb conventional $\ell$-dependent combinatorial normalizations into $Q_A^L$.  In particular, the quadrupole is unit-coupled, $S_{\rm int}^{(2)}=\int\dd\tau\,(Q_{\EE}^{ij}\EE_{ij}+Q_{\BB}^{ij}\BB_{ij})$, matching the convention of Ref.~\cite{Goldberger:2005cd}.

The source-free/minimal action is the sum of the Einstein--Hilbert action $S_{\rm EH}[g]$ for the long-wavelength metric, the minimally coupled point-particle action $S_{\rm pp}[z,g]$, and $S_X[X]$.  Adding the tidal interaction~\eqref{eq:tidal_coupling} gives
\begin{equation}
  S[g,z,X]
  =S_{\rm EH}[g]+S_{\rm pp}[z,g]+S_X[X]+S_{\rm int}[z,g,X].
  \label{eq:full_micro_action}
\end{equation}
When expanding in the tidal coupling, we use the interaction picture with $H_{\rm BH}$ generating the source-free evolution of the black-hole operators and $S_{\rm int}$ treated as the interaction with the external tidal fields.  Standard dissipative worldline EFT matches the retarded multipole correlators, in particular their absorptive parts, to black-hole absorption~\cite{Goldberger:2005cd}; here we extend the same coarse-grained sector to the full CTP so that response and fluctuations are retained together.

\subsection{Closed-time-path influence functional and Gaussian reduced dynamics}
\label{subsec:ctp}

\begin{figure}
\centering
\begin{tikzpicture}[x=1.0cm,y=1.0cm,>=Latex,font=\small]
  \coordinate (Li) at (0,0.75);
  \coordinate (Rf) at (7,0.75);
  \coordinate (Rb) at (7,-0.75);
  \coordinate (Lb) at (0,-0.75);
  \draw[->,thick] (Li) -- (Rf);
  \draw[->,thick] (Rb) -- (Lb);
  \draw[thick] (Rf) -- (Rb);
  \draw[dashed] (Li) -- (Lb);
  \node[above] at (3.5,0.75) {branch 1: forward evolution $U_1$};
  \node[below] at (3.5,-0.75) {branch 2: backward evolution $U_2^\dagger$};
  \node[left,align=right] at (-0.05,0) {$\rho_{\rm tot}^{\rm init}=\rho_{\rm ext}^{\rm init}\otimes\rho_{\rm BH}^{\rm init}$\\factorized initial state};
  \node[right,align=left] at (7.05,0) {$X_1(\tau_f)=X_2(\tau_f)=X_f$\\trace over $X_f$};
  \node[above left] at (Li) {$\tau_i$};
  \node[above right] at (Rf) {$\tau_f$};
\end{tikzpicture}
\caption{CTP contour for the unresolved black-hole variables.  The initial state is~\eqref{eq:initial_factorization}; sewing and integrating the two histories at $\tau_f$ implements $\Tr_{\rm BH}$.}
\label{fig:sk_contour}
\end{figure}
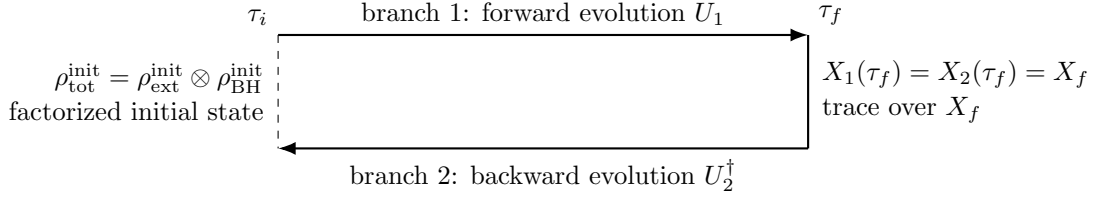

To obtain causal expectation values and the reduced dynamics of the resolved sector, we integrate out $X$ in the in-in/CTP formalism~\cite{Schwinger:1960qe,Keldysh:1964ud,Feynman:1963fq,Kamenev:2011}.  At the initial time $\tau_i$ we assume
\begin{equation}
  \rho_{\rm tot}^{\rm init}
  =\rho_{\rm ext}^{\rm init}\otimes\rho_{\rm BH}^{\rm init},
  \label{eq:initial_factorization}
\end{equation}
where ``ext'' denotes all resolved long-wavelength degrees of freedom and ``BH'' the unresolved $X$ sector.  The exterior state may be pure or mixed; it supplies the initial condition for the subsequent reduced evolution but does not enter the black-hole kernels defined below.  Holding the doubled exterior histories fixed and tracing over the black-hole sector defines
\begin{align}
  \ee^{\ii S_{\IF}[\EE_1,\EE_2,\BB_1,\BB_2]}
  \equiv
  \Tr_{\rm BH}\Bigl[
  U_1[\EE_1,\BB_1] \,\rho_{\rm BH}^{\rm init}\,
  U_2^{\dagger}[\EE_2,\BB_2]
  \Bigr].
  \label{eq:influence_def}
\end{align}
Equivalently, the CTP path integral doubles the unresolved variables, weights their initial endpoints by $\rho_{\rm BH}^{\rm init}$, and sews $X_1(\tau_f)=X_2(\tau_f)$ before tracing over the common final value.  Here $\mathcal X_a$ collectively denotes the electric- and magnetic-type tidal sources on branch $a=1,2$; the contour is shown in Figure~\ref{fig:sk_contour}.

If the exterior degrees of freedom are also integrated over, their two histories are sewn at the final time as well.  For a thermal black-hole state, $\rho_\beta$ may instead be prepared by attaching an imaginary-time segment of length $\beta$ to the contour~\cite{Kamenev:2011}.

Microscopic unitarity further constrains the influence functional.  When the
two exterior histories coincide, the forward and backward evolutions cancel
inside the trace, implying
\begin{equation}
  S_{\IF}[\EE,\EE,\BB,\BB]=0,
  \label{eq:SK_unitarity}
\end{equation}
together with the usual SK reality relation under exchange of the two branches.

Introduce average/difference variables
\begin{equation}
  \EE_{r,L}=\frac{\EE_{1,L}+\EE_{2,L}}{2},
  \qquad
  \EE_{a,L}=\EE_{1,L}-\EE_{2,L},
  \label{eq:ra_vars}
\end{equation}
and analogously for $\BB$, $z^\mu$ and the resolved perturbation $h_{\mu\nu}$ in~\eqref{eq:background_split}.  The physical limit identifies the two CTP histories, so all $a$-variables vanish and the corresponding $r$-variables reduce to the physical fields.  The coupled classical metric--worldline equations are obtained by varying with respect to $h_a$ and $z_a$ before taking this limit, as in Section~\ref{subsec:recoil}.

At Gaussian order in the cumulant expansion in the tidal coupling, the influence action is determined by the source-free Wightman functions
\begin{align}
  S^{>,AB}_{LL'}(\tau,\tau')
  &\equiv\avg{Q_A^L(\tau)Q_B^{L'}(\tau')}_{\rho_{\rm BH}^{\rm init}},\qquad
  S^{<,AB}_{LL'}(\tau,\tau')
  \equiv\avg{Q_B^{L'}(\tau')Q_A^L(\tau)}_{\rho_{\rm BH}^{\rm init}}.
  \label{eq:wightman_time}
\end{align}
It is convenient to use their retarded and symmetric combinations,
\begin{align}
  \DR^{AB}{}_{LL'}(\tau,\tau')
  &\equiv
  \ii\,\theta(\tau-\tau')
  \avg{[Q_A^L(\tau),Q_B^{L'}(\tau')]}_{\rho_{\rm BH}^{\rm init}},
  \label{eq:DRdef}
  \\
  \NN^{AB}{}_{LL'}(\tau,\tau')
  &\equiv
  \frac12\avg{\{Q_A^L(\tau),Q_B^{L'}(\tau')\}}_{\rho_{\rm BH}^{\rm init}}.
  \label{eq:Ndef}
\end{align}
Here $\theta$ is the Heaviside step function.  Stationarity of $\rho_{\rm BH}^{\rm init}$ under $H_{\rm BH}$ makes the kernels functions of $\tau-\tau'$; otherwise they retain two-time dependence.  We use the convention $+\ii\theta[Q,Q]$ for $\DR$.

Expanding~\eqref{eq:influence_def} to second order in the tidal sources gives
\begin{equation}
 S_{\IF}^{(2)}=
 \sum_{A,B,L,L'}\int \dd\tau\dd\tau'\,
 \mathcal{X}^{A,L}_{a}(\tau)
 \left[
 \DR^{AB}{}_{LL'}(\tau,\tau')\mathcal{X}^{B,L'}_{r}(\tau')
 +\frac{\ii}{2}\NN^{AB}_{LL'}(\tau,\tau')\mathcal{X}^{B,L'}_{a}(\tau')
 \right],
 \label{eq:quadratic_IF}
\end{equation}
where $\mathcal{X}^{\EE,L}\equiv\EE^L$ and $\mathcal{X}^{\BB,L}\equiv\BB^L$.  The retarded term gives causal response and the imaginary noise term suppresses differing histories.  Appendix~\ref{app:influence} records the short derivation and our conventions.  Adding the doubled Einstein--Hilbert and point-particle actions gives the full SK effective action.

\subsection{Response, dissipation and noise as real-time Wilson data}
\label{subsec:wilson_data}

For readability, from this point onward we occasionally use a reduced notation when the channel and multipole structure is not essential.  Thus $\DR$ and $\NN$ denote either the full matrices in electric/magnetic--STF-multipole space or a specified diagonal channel, and time-domain products imply the corresponding convolutions.  Explicit sums and indices are retained whenever they affect normalization.

Varying the influence action with respect to $\EE_a$ and then setting the difference sources to zero, a weak physical $r$-source induces
\begin{equation}
  \avg{Q_{\EE}^L(\tau)}_{\mathcal X_r}
  =\int \dd\tau'\,
  \DR^{\EE\EE}{}^{L}{}_{L'}(\tau,\tau')\,
  \EE_r^{L'}(\tau')+\cdots.
  \label{eq:linear_response}
\end{equation}
Here $\avg{\cdots}_{\mathcal X_r}$ denotes the expectation value in the black-hole state driven from $\rho_{\rm BH}^{\rm init}$ by the physical tidal history $\mathcal X_r$ through~\eqref{eq:tidal_coupling}; it reduces to $\avg{\cdots}_{\rho_{\rm BH}^{\rm init}}$ for $\mathcal X_r=0$.  Equation~\eqref{eq:linear_response} is the linear-response relation for the induced multipole, with $\DR$ the causal tidal susceptibility.

The retarded kernel is causal, $\DR(\tau,\tau')=0$ for $\tau<\tau'$.  For a stationary state its Fourier transform $\DR(\omega)$ is the frequency-domain linear-response function.  Using the convention in Equation~\eqref{eq:fourier_convention_intro}, and assuming that no additional infrared time scale obstructs a local derivative expansion,
\begin{equation}
  \DR(\omega)=c_0+c_2\omega^2+\cdots
  +\ii\left(\gamma_1\omega+\gamma_3\omega^3+\cdots\right),
  \label{eq:retarded_expansion}
\end{equation}
where the real analytic coefficients encode conservative finite-size response and the odd imaginary coefficients encode dissipation.  For a Schwarzschild black hole and a fixed multipole channel, matching to low-frequency absorption gives the leading absorptive term~\cite{Goldberger:2005cd}
\begin{equation}
  \operatorname{Im}\DR{}_{A}^{(\ell)}(\omega)
  =\gamma_{A\ell}\,\omega+
  \ord(\omega^3),
  \qquad
  \gamma_{A\ell}\sim r_h^{2\ell+2},
  \label{eq:ohmic_response}
\end{equation}
where the numerical coefficient depends on the normalization of $Q_A^L$ and the tidal tensors.  In an adiabatic inspiral a finite truncation of Equation~\eqref{eq:retarded_expansion} maps onto local PN operators; it does not by itself determine hereditary tails, which require additional low-frequency spectral structure such as poles, thresholds, branch cuts or parametrically long-lived modes.  Modes that remain light on the resolved time scale should instead be retained explicitly in the EFT.

The symmetric kernel $\NN$ is independent fluctuation data unless a state condition relates it to $\DR$.  It controls stochastic multipole fluctuations and, for two distinct histories, decoherence.  Microscopically, however, $\NN$ depends on the occupation of black-hole states and on the same multipole matrix elements that enter the Wightman functions and hence $\DR$; Section~\ref{subsec:eth} makes this common spectral origin explicit.  Thus $\{\DR,\NN,\text{ higher SK cumulants}\}$ may be viewed as generalized real-time Wilson data: absorption fixes the absorptive response, equilibrium KMS relates response and noise, while away from equilibrium the fluctuation data must be matched separately.

The noise kernel is real and symmetric, $\NN^{AB}_{LL'}(\tau,\tau')=\NN^{BA}_{L'L}(\tau',\tau)$.  For any real test function $f_A^L(\tau)$ it satisfies
\begin{equation}
 \int \dd\tau\dd\tau'\,
 f_A^L(\tau)\NN^{AB}_{LL'}(\tau,\tau')f_B^{L'}(\tau')\geq 0,
 \label{eq:N_positive}
\end{equation}
provided it is defined as the covariance of Hermitian operators.  Consequently, in the compact notation introduced above,
\begin{equation}
  \left|\ee^{\ii S_{\IF}^{(2)}}\right|
  =\exp\!\left[-\frac12\mathcal X_a\NN\mathcal X_a\right]\leq1,
  \label{eq:IF_modulus}
\end{equation}
which is the open-system origin of decoherence in the Gaussian approximation.

\subsection{Spectral density, positivity and eigenstate thermalization}
\label{subsec:eth}

The retarded and noise kernels admit a direct microscopic spectral interpretation.  For one Hermitian multipole channel, define the commutator spectral function
\begin{equation}
  \rho_Q(\omega)\equiv S^>(\omega)-S^<(\omega)
  =2\,\operatorname{Im}\DR(\omega).
  \label{eq:spectral_rho}
\end{equation}
Thus classical absorption determines the smooth low-frequency part of the same spectral function that, microscopically, is built from transition matrix elements between black-hole states.  Appendix~\ref{app:spectral} gives the exact spectral decomposition.

\label{subsec:positivity}
Stationarity alone gives a state-independent constraint on noise and absorptive response, even when no fluctuation--dissipation theorem applies.  Define the Fourier-space Wightman matrices
\begin{align}
  S^{>,AB}_{LL'}(\omega)
  &=\int\dd s\,\ee^{\ii\omega s}
  \avg{Q_A^L(s)Q_B^{L'}(0)}_{\rho_{\rm BH}^{\rm init}},
  \quad
  S^{<,AB}_{LL'}(\omega)
  =\int\dd s\,\ee^{\ii\omega s}
  \avg{Q_B^{L'}(0)Q_A^L(s)}_{\rho_{\rm BH}^{\rm init}}.
  \label{eq:wightman_fourier}
\end{align}
Here $S^>(\omega)$ and $S^<(\omega)$ are the Fourier transforms of Eqs.~\eqref{eq:wightman_time}, with the Fourier convention of Equation~\eqref{eq:fourier_convention_intro}.  For a positive stationary density operator in this reference sector, the Wightman functions define positive-semidefinite spectral measures in the combined $(A,L)$ space; wherever ordinary spectral densities exist, $v^\dagger S^{>,<}(\omega)v\geq0$ for any complex channel vector $v$.  Using the reduced matrix notation introduced in Section~\ref{subsec:wilson_data}, one then has
\begin{align}
  \NN(\omega)&=\frac12\left[S^>(\omega)+S^<(\omega)\right],
  \qquad
  \operatorname{Im}\DR(\omega)=\frac12\left[S^>(\omega)-S^<(\omega)\right].
  \label{eq:NDR_wightman}
\end{align}
In these matrix formulas, $\operatorname{Im}\DR\equiv(\DR-\DR^\dagger)/(2\ii)$ denotes the Hermitian absorptive part at fixed real frequency; it reduces to the ordinary imaginary part in a diagonal channel.  The two operator orderings can therefore be reconstructed separately from response and noise:
\begin{equation}
  \NN(\omega)+\operatorname{Im}\DR(\omega)=S^>(\omega)\succeq0,
  \qquad
  \NN(\omega)-\operatorname{Im}\DR(\omega)=S^<(\omega)\succeq0.
  \label{eq:positivity_bound}
\end{equation}
Here $\succeq0$ denotes spectral-measure positivity, reducing to the usual pointwise condition for ordinary densities.  Equation~\eqref{eq:positivity_bound} holds for any stationary state in this zero-mean sector and constrains, but does not determine, $\NN$ and $\DR$.  For a scalar diagonal channel with a spectral density,
\begin{equation}
  \NN_A(\omega)\geq
  \left|\operatorname{Im}\DR{}_{A}(\omega)\right|.
  \label{eq:scalar_bound}
\end{equation}
This standard Wightman-positivity bound, expressed in the black-hole multipole Wilson basis, is weaker than thermal FDT and shows that inelastic response requires a minimum amount of quantum noise~\cite{Biggs:2024doh}.  A stationary nonthermal model must satisfy it while matching its noise independently; a passive zero-temperature channel saturates the bound at nonzero frequency.

\paragraph{Relation to eigenstate thermalization.}
To connect the worldline kernels with microscopic spectral data, consider a regulated unresolved sector with densely spaced, discrete energy levels and coarse-grained microcanonical entropy $\mathcal S(E)$.\footnote{For a narrow energy window of fixed width $\Delta E$, $e^{\mathcal S(E)}\simeq\rho(E)\Delta E$, where $\rho(E)$ is the smoothed density of states.
The mean spacing is $\delta E\simeq\rho(E)^{-1}\simeq\Delta E\,e^{-\mathcal S(E)}$; see Ref.~\cite{Balasubramanian:2006}.
The exterior semiclassical spectrum need not be discrete.}
If this sector accounts for black-hole microstates, semiclassical state counting gives $\mathcal S(E)\simeq S_{\rm BH}(E)=A_H(E)/(4G)$ at leading order, where $A_H(E)$ is the horizon area of the corresponding semiclassical black hole~\cite{BrownYork:1993microcanonical}.
This is state-counting input independent of the present kernel matching; below we retain the general $\mathcal S(E)$, allowing for the quantum corrections discussed in Section~\ref{subsec:qg_matching}.

An EFT frequency bin containing many unresolved transitions probes locally averaged spectral weight. For a chaotic microscopic sector, the eigenstate thermalization hypothesis (ETH) provides an ansatz for the matrix elements of a sufficiently simple multipole operator~\cite{Deutsch:1991,Srednicki:1994,Deutsch:2018eth},
\begin{equation}
  \langle m|Q_A^L|n\rangle
  =\overline Q_A^L(\bar E)\,\delta_{mn}
  +e^{-\mathcal S(\bar E)/2}
  f_A^L(\bar E,\omega_{mn})\,R_{mn},
  \qquad
  \bar E\equiv\frac{E_m+E_n}{2},\quad
  \omega_{mn}\equiv E_m-E_n .
  \label{eq:ETH_ansatz}
\end{equation}
Here $\overline Q_A^L$ is the smooth microcanonical expectation value, $f_A^L$ is a smooth envelope, and $R_{mn}$ describes random-matrix-like fluctuations with $\overline{R_{mn}}=0$ and $\overline{|R_{mn}|^2}=1$ under a local spectral average. The factor $e^{-\mathcal S(\bar E)/2}$ encodes the typical off-diagonal suppression in the ETH ansatz.
For the STF multipoles considered here, rotationally invariant microcanonical averaging gives $\overline Q_A^L=0$.

For a fixed multipole channel and $\omega_{mn}\neq0$, define $W_{mn}\equiv e^{-\mathcal S(\bar E)}|f_A^L(\bar E,\omega_{mn})|^2$.  For $m\neq n$, $W_{mn}\simeq\overline{|\langle m|Q_A^L|n\rangle|^2}$, where $\simeq$ denotes equality after this local spectral averaging.  Inserting the ETH ansatz into the pure-point representation in Appendix~\ref{app:spectral} gives
\begin{align}
 \begin{pmatrix}\overline{S^>(\omega)}_{\rm ETH}\\[1mm]\overline{S^<(\omega)}_{\rm ETH}\end{pmatrix}
 &\simeq 2\pi\!\sum_{m\neq n}W_{mn}\,
 \delta(\omega-\omega_{mn})
 \begin{pmatrix}p_n\\[1mm]p_m\end{pmatrix},
 \label{eq:ETH_wightman}\\[-1mm]
 \overline{\NN(\omega)}_{\rm ETH}
 &\simeq \pi\!\sum_{m\neq n}(p_n+p_m)W_{mn}\,
 \delta(\omega-\omega_{mn}),
 \\
 \overline{\operatorname{Im}\DR(\omega)}_{\rm ETH}
 &\simeq \pi\!\sum_{m\neq n}(p_n-p_m)W_{mn}\,
 \delta(\omega-\omega_{mn}).
 \label{eq:ETH_to_SK}
\end{align}
Here $p_n$ are the probability weights defining the stationary reference ensemble of black-hole energy eigenstates,
\begin{equation}
  \rho_{\rm BH}^{\rm init}=\sum_n p_n\,|n\rangle\langle n|,
  \qquad p_n\geq0,\qquad \sum_n p_n=1.
\end{equation}
The combinations $p_n+p_m$ and $p_n-p_m$ enter the noise
and absorptive response, respectively.

Choosing Gibbs weights, $p_n=Z^{-1}e^{-\beta E_n}$, gives the canonical ensemble, for which detailed balance follows directly from these weights without invoking ETH.
ETH instead describes how coarse-grained correlators of typical high-energy eigenstates or narrow microcanonical windows can approach thermal correlators and approximately reproduce the same detailed-balance structure.  Section~\ref{subsec:kms_fdt} imposes KMS directly on a Gibbs reference state, where the equilibrium relation is exact.  Two-point smoothness alone does not establish quantum chaos; finer spectral correlations and OTOCs probe additional information beyond this matching, as discussed in Section~\ref{subsec:chaos_info}.

\subsection{Hubbard--Stratonovich representation}
\label{subsec:HS}

The standard HS transformation represents the imaginary Gaussian term by an auxiliary classical Gaussian field $\xi_A^L$~\cite{Stratonovich:1957,Hubbard:1959,Kamenev:2011}.  Its labels inherit the tidal/STF structure of $Q_A^L$, but $\xi_A^L$ is an integration variable rather than a microscopic black-hole multipole~\cite{Feynman:1963fq,Hu:2008rga}.  In compact notation,
\begin{equation}
  \exp\!\left[-\frac12\mathcal X_a\NN\mathcal X_a\right]
  =\mathcal Z_\xi^{-1}\int\mathcal D\xi\,
  \exp\!\left[-\frac12\xi\NN^{-1}\xi+\ii\xi\mathcal X_a\right],
  \label{eq:HS}
\end{equation}
with the time, channel and STF contractions inherited from~\eqref{eq:quadratic_IF}.  The auxiliary field obeys
\begin{equation}
  \avg{\xi_A^L(\tau)}_{\rm st}=0,
  \qquad
  \avg{\xi_A^L(\tau)\xi_B^{L'}(\tau')}_{\rm st}
  =\NN^{AB}_{LL'}(\tau,\tau').
  \label{eq:xi_cov}
\end{equation}
It is a stochastic representation of the same reduced dynamics, not a new microscopic degree of freedom.  Because the same $\xi_A^L$ couples through the tidal field, variation of the resolved metric and worldline produces coupled stochastic Einstein--worldline equations rather than a worldline Langevin force on a fixed geometry.  Section~\ref{subsec:recoil} solves this coupled system realization by realization before stochastic averaging, while Appendix~\ref{app:hs_langevin} gives the covariant source construction, leading-PN reduction and semiclassical scaling.

\section{Thermal equilibrium, fluctuation--dissipation and dynamical KMS}
\label{sec:kms}

\subsection{Microscopic KMS identity and the fluctuation--dissipation theorem}
\label{subsec:kms_fdt}

For a thermal choice of the initial black-hole state,
\begin{equation}
  \rho_{\rm BH}^{\rm init}=\rho_\beta=\frac{e^{-\beta H_{\rm BH}}}{Z(\beta)},
  \label{eq:thermalrho}
\end{equation}
the stationary multipole correlators obey the bosonic KMS identity
\begin{equation}
  \avg{Q_A^L(\tau-\ii\beta)Q_B^{L'}(0)}_{\rho_\beta}
  =\avg{Q_B^{L'}(0)Q_A^L(\tau)}_{\rho_\beta},
  \label{eq:KMS_time}
\end{equation}
which implies detailed balance,
\begin{equation}
  S^<(\omega)=e^{-\beta\omega}S^>(\omega),
  \label{eq:detailed_balance}
\end{equation}
and hence the quantum FDT~\cite{Kubo:1966nj},
\begin{equation}
  \NN(\omega)=\coth\!\left(\frac{\beta\omega}{2}\right)
  \operatorname{Im}\DR(\omega).
  \label{eq:FDT}
\end{equation}
KMS constrains the source-free reference correlators; it does not require the tidally driven state to remain Gibbs.  The Hartle--Hawking state is the canonical stationary example for an eternal nonextremal black hole~\cite{Hartle:1976tp,Kay:1988mu}.  Detailed balance follows directly from Gibbs weights, whereas ETH concerns when coarse-grained eigenstate or microcanonical correlators approach this thermal structure (Section~\ref{subsec:eth}).

For an equilibrium Schwarzschild black hole,
\begin{equation}
  \beta=\beta_H=T_H^{-1},
  \qquad
  T_H=\frac{\kappa}{2\pi}=\frac{1}{8\pi GM}.
  \label{eq:hawkingT}
\end{equation}
The temperature and spectra must be matched together to the equilibrium geometry.\footnote{The normalized real-time influence functional does not by itself determine the absolute black-hole entropy, which requires equilibrium matching. For Schwarzschild, the matched temperature in Eq.~\eqref{eq:hawkingT} and the first law $\dd S_{\rm BH}=\dd M/T_H$ reproduce the Bekenstein--Hawking entropy $S_{\rm BH}=A_H/(4G)$, where $A_H=16\pi G^2M^2$, up to an additive constant fixed by additional thermodynamic input~\cite{Bekenstein:1973ur,Hawking:1975vcx}.}
In a suitable rotating or charged KMS setting the thermal weight uses the horizon-frame frequency $\widetilde\omega=\omega-m\Omega_H-q\Phi_H$, with $m$ and $q$ the azimuthal and electric charges.  Kerr requires a spinning worldline CTP and separately matched response/noise, including superradiant channels~\cite{Goldberger:2020yvj}; this conditional statement does not assume a global Kerr Gibbs state~\cite{Kay:1988mu}.  Neither rotating nor charged extensions are developed here.

Combining FDT with the Ohmic response~\eqref{eq:ohmic_response} gives
\begin{equation}
  \NN_A^{(\ell)}(\omega)
  =\gamma_{A\ell}\omega
  \coth\!\left(\frac{\beta_H\omega}{2}\right)+\cdots.
  \label{eq:thermal_noise_spectrum}
\end{equation}
Along a physical near-extremal family, both $T_H$ and the matched response coefficients vary, so the order of the $\omega\to0$ and $T_H\to0$ limits must be specified.  For $|\omega|\ll T_H$,
\begin{equation}
  \NN_A^{(\ell)}(\omega)
  =2T_H\gamma_{A\ell}
  +\ord\!\left(\frac{\omega^2}{T_H}\right),
  \label{eq:white_noise}
\end{equation}
so the leading noise is approximately white.  This requires the additional hierarchy $|\omega|\ll T_H$; $r_h|\omega|\ll1$ alone is insufficient, especially near extremality.  In the opposite quantum regime $T_H\ll|\omega|$,
\begin{equation}
  \NN_A^{(\ell)}(\omega)
  \longrightarrow
  \operatorname{sgn}(\omega)\operatorname{Im}\DR{}_{A}^{(\ell)}(\omega),
  \label{eq:quantum_noise_limit}
\end{equation}
which scales as $|\omega|$ only while the matched response remains Ohmic.  Near-extremal quantum corrections can modify that spectrum, and noise in a nonthermal state must be obtained from its Wightman data; see Section~\ref{subsec:qg_matching}.

\subsection{Dynamical KMS as the infrared encoding of equilibrium}
\label{subsec:dynkms}

When the microscopic theory also admits the appropriate antiunitary reversibility operation, the KMS property can be represented in a local SK EFT as a discrete dynamical-KMS symmetry~\cite{Sieberer:2015hba,Crossley:2015evo,Glorioso:2017fpd,Haehl:2015pja,Haehl:2015foa,Akyuz:2024coset,Lau:2025gravity}.  In the classical derivative limit a bosonic variable transforms schematically as
\begin{align}
  q_r(\tau)&\rightarrow \Theta q_r(-\tau),
  \qquad
  q_a(\tau)\rightarrow \Theta q_a(-\tau)
  +\ii\beta\,\Theta\partial_\tau q_r(-\tau)+\cdots,
  \label{eq:dynKMSschematic}
\end{align}
where the omitted terms contain higher-derivative quantum corrections.  The logical order is
\begin{equation}
\text{thermal state}\Rightarrow\text{microscopic KMS}\Rightarrow\text{FDT},
\qquad
\text{KMS}+\Theta\text{ invariance}\Rightarrow\text{dynamical KMS},
\label{eq:kms_chain}
\end{equation}
so dynamical KMS is the infrared symmetry representation of the same equilibrium input, not an additional black-hole postulate.  For a generic nonthermal state it should not be imposed.

For one diagonal tidal channel the leading local terms are
\begin{equation}
  S_{\IF,A\ell}^{\rm local}
  =\int\dd\tau\left[
  c_{0,A\ell}\,\mathcal X^{A,L}_{a}\mathcal X^{A}_{r,L}
  -\gamma_{A\ell}\mathcal X^{A,L}_{a}\dot{\mathcal X}^{A}_{r,L}
  +\ii D_{A\ell}\mathcal X^{A,L}_{a}\mathcal X^{A}_{a,L}
  \right]+\cdots,
  \label{eq:local_IF}
\end{equation}
where $c_{0,A\ell}$ encodes the static conservative response and the dot denotes a proper-time derivative.  Dynamical KMS does not fix this conservative coefficient; it relates the dissipative and noise coefficients as
\begin{equation}
  D_{A\ell}=\gamma_{A\ell}T_H,
  \label{eq:Einstein_local}
\end{equation}
in the Schwarzschild thermal regime, with $\NN_A^{(\ell)}(0)=2D_{A\ell}$ in our convention.  For a Schwarzschild black hole in four-dimensional GR, matching gives $c_{0,A\ell}=0$, reflecting the vanishing static tidal Love numbers~\cite{Chia:2020yla}.  Equation~\eqref{eq:Einstein_local} is simply the local $|\omega|\ll T_H$ limit of the full quantum FDT~\eqref{eq:FDT}; the frequency-space relation remains the more general statement.

\subsection{Nonequilibrium states and local equilibrium}
\label{subsec:thermalization}

The general CTP construction does not assume equilibrium.  A stationary but nonthermal density matrix gives time-translation-invariant kernels $\DR(\omega)$ and $\NN(\omega)$ with no KMS relation: their spectra must be specified separately, but they still obey the exact nonthermal inequalities~\eqref{eq:positivity_bound} and~\eqref{eq:scalar_bound}.  Thus stationarity suffices for spectral positivity, while a Gibbs/KMS condition is needed to infer the noise from the absorptive response by FDT~\eqref{eq:FDT}.  Genuine two-time kernels are needed when the reference state or background evolves appreciably.  This distinction matters for collapse and merger remnants, the Unruh state on a fixed black-hole background (stationary but non-KMS), and backgrounds that evolve appreciably due to evaporation or strong driving~\cite{Berti:2009kk,Unruh:1976evap}.  Thermal-looking correlators of selected observables do not by themselves establish a global Gibbs state.

When the central-time evolution is slow, define
\begin{equation}
  F(\bar\tau,\omega)=\int\dd s\,e^{\ii\omega s}
  F\!\left(\bar\tau+\frac{s}{2},\bar\tau-\frac{s}{2}\right),
  \qquad
  \bar\tau=\frac{\tau+\tau'}{2},\quad s=\tau-\tau',
  \label{eq:wigner}
\end{equation}
and expand in $\partial_{\bar\tau}$.  This Wigner/gradient expansion is controlled only when the macroscopic evolution time is much longer than the kernel correlation time.  If the state is also close to local equilibrium, FDT holds at leading adiabatic order with slowly varying parameters; slow geometric evolution alone does not establish local KMS.  Appendix~\ref{app:noneq} gives the corresponding expansion and fluctuation--response diagnostic.

Table~\ref{tab:regimes} summarizes the distinct assumptions and their consequences.  A frequency- and channel-dependent fluctuation--response ratio may serve as a diagnostic of local equilibration, but without local KMS it is not by itself a thermodynamic temperature; its precise definition is given in Appendix~\ref{app:noneq}.

\begin{table}[t]
\centering
\small
\renewcommand{\arraystretch}{1.18}
\begin{tabularx}{\textwidth}{@{}>{\raggedright\arraybackslash}p{0.23\textwidth}>{\raggedright\arraybackslash}p{0.25\textwidth}>{\raggedright\arraybackslash}X@{}}
\toprule
Reference state & Kernel structure & Noise--response statement \\
\midrule
Gibbs/KMS (stationary)
& $\DR(\omega),\ \NN(\omega)$
& Exact FDT fixes $\NN$ from $\operatorname{Im}\DR$; \eqref{eq:FDT}. \\
\midrule
Nonthermal (stationary)
& $\DR(\omega),\ \NN(\omega)$
& Independent spectra obey $\NN\pm\operatorname{Im}\DR\succeq0$; \eqref{eq:positivity_bound}. \\
\midrule
Local equilibrium (slow evolution)
& Wigner kernels $\DR(\bar\tau,\omega),\ \NN(\bar\tau,\omega)$
& Leading-order local FDT, subject to gradient corrections; \eqref{eq:E_localFDT}. \\
\midrule
General nonstationary state
& Two-time kernels $\DR(\tau,\tau'),\ \NN(\tau,\tau')$
& Positive noise kernel; no general FDT; \eqref{eq:N_positive}. \\
\bottomrule
\end{tabularx}
\caption{What follows from stationarity, thermality and slow local equilibration.  The stationary positivity bound holds with or without KMS; the two-time noise covariance is positive even without stationarity.}
\label{tab:regimes}
\end{table}

\section{Applications}
\label{sec:applications}

At Gaussian order, the unresolved black-hole sector enters through the retarded response $\DR$ and symmetric noise kernel $\NN$.  We use the HS representation of $\NN$ for stochastic metric--worldline dynamics in Section~\ref{subsec:recoil}, and the same kernel directly in the two-history influence functional in Section~\ref{subsec:decoherence_general}.  For the nonspinning Schwarzschild electric quadrupole, long-wavelength absorbing matching in the unit-coupled convention of Section~\ref{sec:worldline} gives~\cite{Goldberger:2005cd,Kobayashi:2025tdn}
\begin{equation}
  \operatorname{Im}\DR{}_{\EE}^{(2)}(\omega)
  =\gamma_{\EE2}\omega+\ord(\omega^3),
  \qquad
  \gamma_{\EE2}=\frac{16}{45}G^5M^6,
  \label{eq:common_schw_response}
\end{equation}
Equation~\eqref{eq:common_schw_response} does not assume thermality.  For the equilibrium applications below, we use the Sec.~\ref{subsec:kms_fdt} result, Eq.~\eqref{eq:FDT}, whose low-frequency limit is Eq.~\eqref{eq:white_noise}.  Thus, for a stationary KMS state at $T_H=(8\pi GM)^{-1}$ and $\beta_H|\omega|\ll1$,
\begin{equation}
  \NN_{\EE}^{ij,kl}(0)
  =2T_H\gamma_{\EE2}\,\mathcal P_{\rm STF}^{ij,kl}
  =\frac{4}{45\pi}G^4M^5\,\mathcal P_{\rm STF}^{ij,kl}.
  \label{eq:common_schw_noise}
\end{equation}
Here $\mathcal P_{\rm STF}^{ij,kl}\equiv \frac12(\delta^{ik}\delta^{jl}+\delta^{il}\delta^{jk})-\frac13\delta^{ij}\delta^{kl}$ projects onto rank-two STF tensors in the local spatial frame.  When the dissipative coefficient is fixed by horizon absorption, the following map applies: horizon absorption $\to\operatorname{Im}\DR\to\NN$; only the last step uses KMS/FDT.  In a stationary non-KMS state, $\NN(\omega)$ is instead independent matching data.

\subsection{Thermal noise and coupled stochastic metric--worldline dynamics}
\label{subsec:recoil}

Building on the Hubbard--Stratonovich representation in Eqs.~\eqref{eq:HS} and \eqref{eq:xi_cov}, our goal is to propagate the matched noise kernel into stochastic fluctuations of the resolved classical dynamics.  Because the same realization $\xi_A^L$ sources both resolved fields, the metric $h_{\mu\nu}$ and worldline $z^\mu$ must be evolved as a coupled stochastic system rather than treating the geometry as fixed.  We first formulate this coupled evolution and its projection onto mass and acceleration, and then specialize to a nonspinning black-hole binary to evaluate the leading-PN stochastic force and its covariance.

\subsubsection{Coupled metric--worldline stochastic system}
\label{subsubsec:coupled_stochastic_system}

For a fixed HS realization $\xi_A^L$, the coupled stochastic Einstein--worldline equations are
\begin{equation}
 \left.\frac{\delta S_{\rm det}}{\delta h_{a\mu\nu}(x)}\right|_{a=0}+\mathscr J_\xi^{\mu\nu}(x)=0,
 \qquad
 \left.\frac{\delta S_{\rm det}}{\delta z_{a\mu}(\tau)}\right|_{a=0}+F_\xi^\mu(\tau)=0,
 \label{eq:coupled_metric_worldline_eom}
\end{equation}
where $S_{\rm det}$ omits the HS term but retains the deterministic retarded tidal response.  The corresponding metric source and worldline force are
\begin{equation}
  \mathscr J_\xi^{\mu\nu}\equiv\left.\frac{\delta S_\xi}{\delta h_{a\mu\nu}}\right|_{a=0},
  \qquad
  F_\xi^\mu\equiv\left.\frac{\delta S_\xi}{\delta z_{a\mu}}\right|_{a=0},
  \qquad
  F_\xi^\mu\big|_{u,e}=\sum_{A,L}\xi_A^L\nabla^\mu\mathcal X_L^A .
  \label{eq:covariant_random_force}
\end{equation}
The last expression varies only the position dependence of the tidal pullback at fixed $u^\mu$ and $e_i^\mu$; proper-time, velocity and frame variations are given in Appendix~\ref{app:covariant_coupled_solution}.

Let $(\bar h,\bar z)$ be the deterministic retarded solution at $\xi=0$ with the same initial data and gauge fixing.  Write
\[
 h_r=\bar h+h_\xi,\qquad z_r=\bar z+z_\xi,
\]
where $(h_\xi,z_\xi)$ are the parts linear in the stochastic source.  Linearizing Eq.~\eqref{eq:coupled_metric_worldline_eom} about $(\bar h,\bar z)$ gives
\begin{equation}
 \begin{pmatrix}L_{hh}&L_{hz}\\ L_{zh}&L_{zz}\end{pmatrix}
 \binom{h_\xi}{z_\xi}
 =\binom{B_h}{B_z}\xi .
 \label{eq:linear_coupled_noise_blocks}
\end{equation}
The first subscript on $L$ labels the equation and the second the fluctuating field; e.g. $L_{hz}$ varies the metric equation with respect to the worldline.  All blocks are evaluated on $(\bar h,\bar z)$, with $B_h\xi=-\mathscr J_\xi[\bar h,\bar z]$ and $B_z\xi=-F_\xi[\bar h,\bar z]$.  We gauge-fix $L_{hh}$; Appendix~\ref{app:covariant_coupled_solution} uses de Donder gauge.

The tidal perturbation from $h_\xi$ is already contained in $L_{zh}h_\xi$ and is not a second random force.  Terms such as $\xi\,\delta\mathcal X[h_\xi]$ are quadratic in the stochastic amplitude and are omitted.  Equation~\eqref{eq:linear_coupled_noise_blocks} is solved retardedly for each $\xi$ before averaging.

Let $G_h^R\equiv(L_{hh})^{-1}_{\rm ret}$ be the gauge-fixed retarded metric propagator.  Eliminating $h_\xi$ gives
\begin{equation}
 \begin{gathered}
   L_{\rm eff}z_\xi=B_{\rm eff}\xi,\qquad
   L_{\rm eff}=L_{zz}-L_{zh}G_h^RL_{hz},\qquad
   B_{\rm eff}=B_z-L_{zh}G_h^RB_h,\\
   G_{z,\rm eff}^R\equiv(L_{\rm eff})^{-1}_{\rm ret},\qquad
   z_\xi=T_z\xi,\quad h_\xi=T_h\xi,\qquad
   T_z=G_{z,\rm eff}^RB_{\rm eff},\quad
   T_h=G_h^R(B_h-L_{hz}T_z).
 \end{gathered}
 \label{eq:noise_transfer_main}
\end{equation}
Here $-L_{zh}G_h^RL_{hz}$ is retarded metric feedback; in the binary application below it dresses the orbital propagator.  By contrast, $-L_{zh}G_h^RB_h$ is the metric-mediated part of the same noise that acts directly through $B_z\xi$.  The stochastic metric is therefore feedback, not an independent noise source.

Only after this fixed-realization solution do we form the connected Gaussian covariances,
\begin{equation}
 C_{zz}=T_z\NN T_z^\dagger,\qquad
 C_{hh}=T_h\NN T_h^\dagger,\qquad
 C_{hz}=T_h\NN T_z^\dagger .
 \label{eq:joint_noise_covariance_main}
\end{equation}
For example, $C_{zz}\equiv\avg{z_\xi z_\xi^\dagger}_{\rm st}$, with time and index contractions implicit.  Direct and metric-mediated forcing interfere coherently because they share the same $\xi$; $C_{hz}\neq0$ correlates metric/waveform and orbital perturbations.  Propagation through $T_h$ and $T_z$ generally colors even white source noise.

After substituting the retarded metric solution used in Eq.~\eqref{eq:noise_transfer_main}, the worldline equation in Eq.~\eqref{eq:coupled_metric_worldline_eom} reduces at linear order to $L_{\rm eff}z_\xi=B_{\rm eff}\xi$.  With $p^\mu\equiv M u^\mu$ and $a^\mu\equiv Du^\mu/\dd\tau$, its temporal and spatial projections are
\begin{equation}
\frac{Dp^\mu}{d\tau}=F^\mu_{\rm eff}[h_r,z_r;\xi],
\qquad
\dot M=-u_\mu F^\mu_{\rm eff},
\qquad
M a^\mu=P^\mu{}_{\nu}F^\nu_{\rm eff},
\label{eq:mass_acceleration_decomposition}
\end{equation}
where a dot denotes $\dd/\dd\tau$: the parallel component changes the mass and the orthogonal component the motion.  The deterministic part gives the usual tidal heating.  For diagonal Schwarzschild channels,
\begin{equation}
  \Delta M_{\rm diss}
  =\sum_{A,\ell}\int\frac{\dd\omega}{2\pi}\,
  \omega\,\bar{\mathcal X}^{A,L}(-\omega)
  \operatorname{Im}\DR{}_{A}^{(\ell)}(\omega)
  \bar{\mathcal X}^{A}_{L}(\omega)+\cdots,
  \label{eq:mass_absorption_balance}
\end{equation}
with $\bar{\mathcal X}$ evaluated on $(\bar h,\bar z)$.  The stochastic mass rate is another projection, $\delta\dot M_\xi=T_M\xi$, with covariance $T_M\NN T_M^\dagger$; its direct term begins as $\sum_{A,L}\xi_A^L\dot{\bar{\mathcal X}}{}^A_L$, while metric- and worldline-mediated pieces enter through $h_\xi$ and $z_\xi$.  Orbit, metric and localized energy transfer are therefore correlated observables of the same noise-driven system.

\subsubsection{Leading-PN stochastic force in binary black-hole dynamics}
\label{subsubsec:binary_stochastic_force}

For an explicit acceleration estimate, consider a PN binary of two nonspinning Schwarzschild black holes.  Let $I,J\in\{1,2\}$ label the bodies, $\bm r=\bm z_1-\bm z_2$, $r=|\bm r|$, $n^i=r^i/r$, and $M_{\rm tot}=M_1+M_2$.  Retaining only the electric quadrupole, we assign each body its low-frequency Hartle--Hawking KMS kernel and take the two noises to be independent,
\begin{equation}
 \avg{\xi_I^{ij}(t)\xi_J^{kl}(t')}_{\rm st}
 =\delta_{IJ}\,\frac{4}{45\pi}G^4M_I^5\,
 \mathcal P_{\rm STF}^{ij,kl}\,\delta(t-t').
 \label{eq:binary_noise_covariance}
\end{equation}
A non-KMS state instead requires independently matched kernels $\NN_I(\omega)$.

At the first nonvanishing PN order, potential-region counting makes instantaneous electric-quadrupole exchange the unique finite inter-body term linear in $\xi$.  Integrating out the Newtonian potential combines $B_z\xi$ with the finite inter-body part of $-L_{zh}G_h^RB_h\xi$; coincident propagation is excluded from this finite inter-body contribution.  Retardation and 1PN electric/magnetic effects enter at relative $\ord(v^2)$, with radiation and tails further suppressed.  Appendix~\ref{app:pn_coupled_noise} discusses these PN corrections in detail.

For body $I$, let $a_{I,\xi}^i$ denote the part of its PN coordinate acceleration $\ddot z_I^i$ that is linear in the horizon noise.  This is distinct from the covariant four-acceleration $a^\mu$ in Eq.~\eqref{eq:mass_acceleration_decomposition}.  The stochastic relative acceleration is
\begin{equation}
 a_\xi^i\equiv a_{1,\xi}^i-a_{2,\xi}^i=\ddot r_\xi^i.
\end{equation}
At leading PN order, its covariance is
\begin{equation}
\begin{gathered}
  \avg{a_{\xi,k}(t)a_{\xi,l}(t')}_{\rm st}
  =C_0^{\rm full}(\delta_{kl}+2n_kn_l)\delta(t-t'),\\
  C_0^{\rm full}
  =\frac{8}{5\pi}\frac{G^6M_{\rm tot}^2(M_1^3+M_2^3)}{r^8}.
\end{gathered}
\label{eq:main_feedback_full}
\end{equation}
Here ``full'' means complete for the relative-acceleration source at this leading PN order, including the coherent direct and metric-mediated terms.  Translational invariance gives equal-and-opposite stochastic forces realization by realization.  It is not the all-orders orbital covariance $C_{zz}$, which also receives dressed-propagator, self-force, radiation and tail effects; Appendix~\ref{app:covariant_coupled_solution} gives the fixed-metric comparison.

At the Gaussian-linear order kept here, $\avg{\xi}_{\rm st}=0$, so $\NN$ does not shift the mean inspiral trajectory; its first orbital imprint is the acceleration covariance and the phase diffusion generated by it.  Restoring $\hbar$ makes the quantum suppression explicit,
\begin{equation}
  C_0^{\rm full}=\frac{8}{5\pi}\ell_{\rm P}^2
  \frac{r_{\rm tot}^2(r_1^3+r_2^3)}{r^8},
  \qquad r_I\equiv GM_I,\quad r_{\rm tot}\equiv GM_{\rm tot},
  \label{eq:main_feedback_planck}
\end{equation}
so the acceleration variance is $\ord(\ell_{\rm P}^2)$ and the rms amplitude $\ord(\ell_{\rm P})$.  Metric feedback changes the coefficient but not this Planck suppression.  Appendix~\ref{app:pn_coupled_noise} further shows that the associated secular phase diffusion remains Planck suppressed for $\Omega^{-1}\ll t\ll t_{\rm rr}$ in macroscopic binaries.

The same two-body dynamics also generates mass fluctuations.  For body $I$ tidally driven by $K\neq I$, the leading direct term is $\delta\dot M_{I,\xi}\simeq\xi_I^{ij}\dot{\bar{\mathcal E}}{}^{(K)}_{I,ij}$.  If the two black-hole sectors are initially independent, the leading body-label covariance is therefore a self-correlation; cross-body correlations require mixing or correlated internal states.  The same local tidal vertex and state-dependent kernel $\NN_I$ also determine the mass-shift variance in an unbound encounter.  As a separate check of this noise matching, Appendix~\ref{app:scattering_mass_variance} evaluates the mass projection on the leading straight-line scattering trajectory relevant to Ref.~\cite{Aoki:2025quantum} and reproduces the nonrelativistic Unruh and Boulware coefficients.\footnote{\label{fn:vacuum_choice}In Schwarzschild, the Hartle--Hawking state is thermal in both the horizon and incoming sectors.  For the Unruh state, the positive-frequency part agrees with Hartle--Hawking, while the incoming vacuum contribution has support only through $\theta(\omega)$, so the corresponding negative-frequency fluctuations are absent~\cite{Goldberger:2019sya,Unruh:1976evap}.  The Boulware state is recovered as the zero-temperature limit of the matched stationary correlator~\cite{Boulware:1975Schwarzschild}; see Footnote~\ref{fn:lowT_vacuum} in Sec.~\ref{subsec:infrared_noise_decoh}.  State dependence resides in the Wightman correlators, and hence in $\NN$, not in the local tidal vertex.}  This comparison uses the same scattering geometry; it does not identify the bound-orbit covariance above with the scattering observable.

\subsection{DSW horizon decoherence from the two-history influence functional}
\label{subsec:decoherence_general}

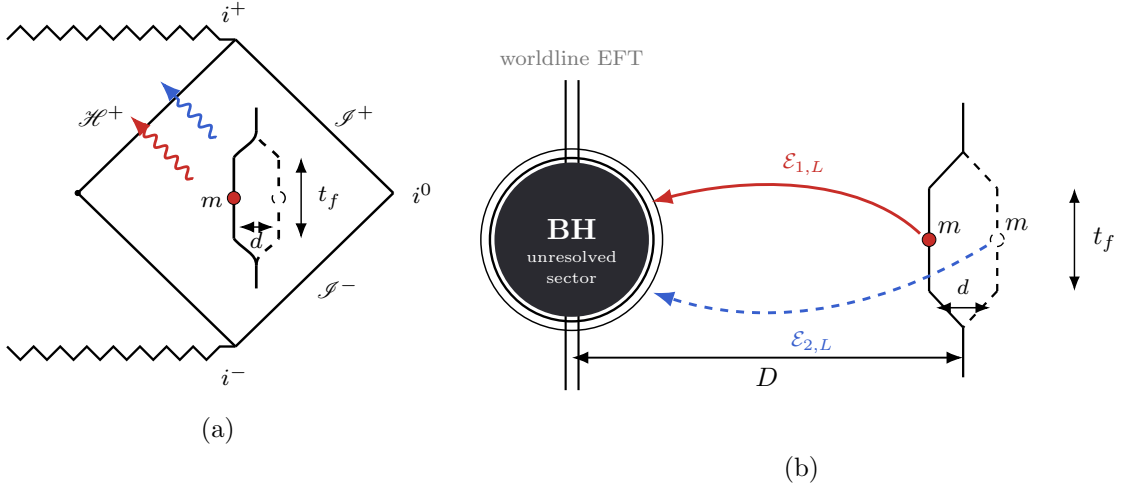
\begin{figure}
\centering
\begin{subfigure}[t]{0.41\textwidth}
  \centering
  \resizebox{\linewidth}{!}{
\begingroup
\definecolor{branchone}{RGB}{200,45,40}
\definecolor{branchtwo}{RGB}{55,95,205}
\begin{tikzpicture}[x=1cm,y=1cm,>=Latex,
  every node/.style={font=\small},
  waveA/.style={branchtwo, line width=0.95pt},
  waveB/.style={branchone, line width=0.95pt}]

\coordinate (B)  at (1.05,2.25);
\coordinate (ip) at (2.95,4.10);
\coordinate (im) at (2.95,0.40);
\coordinate (i0) at (4.85,2.25);
\coordinate (cutT) at (0.20,4.10);
\coordinate (cutB) at (0.20,0.40);

\draw[line width=0.85pt] (ip)--(i0);
\draw[line width=0.85pt] (i0)--(im);
\draw[line width=0.85pt] (B)--(ip);
\draw[line width=0.85pt] (B)--(im);

\draw[line width=0.80pt]
  (cutT)--(0.36,4.26)--(0.52,4.10)--(0.68,4.26)--(0.84,4.10)--(1.00,4.26)--
  (1.16,4.10)--(1.32,4.26)--(1.48,4.10)--(1.64,4.26)--(1.80,4.10)--
  (1.96,4.26)--(2.12,4.10)--(2.28,4.26)--(2.44,4.10)--(2.60,4.26)--(2.76,4.10)--(ip);
\draw[line width=0.80pt]
  (cutB)--(0.36,0.24)--(0.52,0.40)--(0.68,0.24)--(0.84,0.40)--(1.00,0.24)--
  (1.16,0.40)--(1.32,0.24)--(1.48,0.40)--(1.64,0.24)--(1.80,0.40)--
  (1.96,0.24)--(2.12,0.40)--(2.28,0.24)--(2.44,0.40)--(2.60,0.24)--(2.76,0.40)--(im);

\node[anchor=west, font=\scriptsize] at (4.94,2.25) {$i^0$};
\node[anchor=south, font=\scriptsize] at (2.94,4.18) {$i^+$};
\node[anchor=north, font=\scriptsize] at (2.94,0.30) {$i^-$};
\node[font=\scriptsize,] at (1.34,3.18) {$\mathscr H^+$};
\node[font=\scriptsize] at (4.38,3.18) {$\mathscr I^+$};
\node[font=\scriptsize] at (4.18,1.10) {$\mathscr I^-$};
\filldraw[black] (B) circle (0.030);

\draw[line width=0.92pt] (3.20,1.10) -- (3.20,1.38);
\draw[line width=0.92pt] (3.20,1.38) .. controls (3.20,1.52) and (3.05,1.56) .. (2.93,1.70);
\draw[dashed, line width=0.92pt] (3.20,1.38) .. controls (3.20,1.52) and (3.35,1.56) .. (3.47,1.70);
\draw[line width=0.92pt] (2.93,1.70) -- (2.93,2.68);
\draw[dashed, line width=0.92pt] (3.47,1.70) -- (3.47,2.68);
\draw[line width=0.92pt] (2.93,2.68) .. controls (3.05,2.82) and (3.20,2.86) .. (3.20,3.00);
\draw[dashed, line width=0.92pt] (3.47,2.68) .. controls (3.35,2.82) and (3.20,2.86) .. (3.20,3.00);
\draw[line width=0.92pt] (3.20,3.00) -- (3.20,3.28);

\filldraw[branchone, draw=black, line width=0.25pt] (2.93,2.19) circle (0.078);
\filldraw[white, draw=black, dashed, line width=0.50pt] (3.47,2.19) circle (0.078);
\node[font=\scriptsize, anchor=east] at (2.93,2.14) {$m$};
\draw[<->, line width=0.50pt] (2.99,1.84)--(3.41,1.84);
\node[font=\scriptsize] at (3.20,1.69) {$d$};
\draw[<->, line width=0.50pt] (3.74,1.70)--(3.74,2.68);
\node[font=\scriptsize, anchor=west] at (3.81,2.19) {$t_f$};

\draw[waveA, decorate, decoration={snake,amplitude=0.055cm,segment length=0.18cm}]  (2.72,2.92) -- (2.21,3.43);
\draw[waveA, <-, shorten <=0pt, shorten >=0pt] (2.04,3.60) -- (2.11,3.53);
\draw[waveB, decorate, decoration={snake,amplitude=0.055cm,segment length=0.18cm}]  (2.44,2.42) -- (1.75,3.11);
\draw[waveB, <-, shorten <=0pt, shorten >=0pt]  (1.68,3.18) -- (1.75,3.11);

\end{tikzpicture}
\endgroup}
  \caption{}
\end{subfigure}\hfill
\begin{subfigure}[t]{0.57\textwidth}
  \centering
\begingroup
\definecolor{bhblack}{RGB}{42,42,48}
\definecolor{branchone}{RGB}{200,45,40}
\definecolor{branchtwo}{RGB}{55,95,205}
\definecolor{softgray}{RGB}{120,120,120}
\begin{tikzpicture}[x=1cm,y=1cm,>=Latex,
  every node/.style={font=\small},
  tideA/.style={branchone, line width=1.10pt},
  tideB/.style={branchtwo, dashed, line width=1.10pt}]
\draw[line width=0.8pt] (0.90,0.35)--(0.90,4.45);
\draw[line width=0.8pt] (1.07,0.35)--(1.07,4.45);
\node[text=softgray, font=\scriptsize, anchor=south] at (0.98,4.52) {worldline EFT};
\draw[line width=0.90pt] (0.98,2.34) circle (1.08);
\draw[line width=0.55pt] (0.98,2.34) circle (1.20);
\fill[bhblack] (0.98,2.34) circle (1.02);
\node[text=white, font=\bfseries] at (0.98,2.48) {BH};
\node[text=white, font=\tiny, align=center] at (0.98,1.98) {unresolved\\sector};
\draw[line width=0.85pt] (6.15,4.15) -- (6.15,3.50);
\draw[line width=0.85pt] (6.15,3.50) -- (5.70,3.02);
\draw[dashed, line width=0.85pt] (6.15,3.50) -- (6.60,3.02);
\draw[line width=0.85pt] (5.70,3.02) -- (5.70,1.66);
\draw[dashed, line width=0.85pt] (6.60,3.02) -- (6.60,1.66);
\draw[line width=0.85pt] (5.70,1.66) -- (6.15,1.18);
\draw[dashed, line width=0.85pt] (6.60,1.66) -- (6.15,1.18);
\draw[line width=0.85pt] (6.15,1.18) -- (6.15,0.52);
\filldraw[branchone, draw=black, line width=0.25pt] (5.70,2.34) circle (0.09);
\filldraw[white, draw=black, dashed, line width=0.50pt] (6.60,2.34) circle (0.09);
\node[anchor=west] at (5.68,2.50) {$m$};
\node[anchor=west] at (6.57,2.54) {$m$};
\draw[<->, line width=0.58pt] (5.79,1.44)--(6.51,1.44);
\node[font=\scriptsize, anchor=south] at (6.15,1.49) {$d$};
\draw[<->, line width=0.55pt] (7.62,1.66)--(7.62,3.02);
\node[anchor=west] at (7.73,2.34) {$t_f$};
\draw[<->, line width=0.72pt] (0.98,0.78)--(6.15,0.78);
\node at (3.56,0.52) {$D$};
\draw[tideA, ->] (5.58,2.42) .. controls (4.75,3.20) and (3.25,3.16) .. (2.03,2.84);
\draw[tideB, ->] (6.46,2.26) .. controls (4.85,1.28) and (3.35,1.22) .. (2.05,1.64);
\node[branchone, font=\scriptsize] at (4.05,3.34) {$\mathcal E_{1,L}$};
\node[branchtwo, font=\scriptsize] at (4.16,0.98) {$\mathcal E_{2,L}$};
\end{tikzpicture}
\endgroup
  \caption{}
\end{subfigure}
\caption{DSW setup in the Schwinger--Keldysh worldline EFT. (a) Full black-hole spacetime: a mass $m$ is held in a spatial superposition for a time $t_f$, sourcing branch-dependent tidal fields. (b) Coarse-grained description: the black hole is replaced by an unresolved worldline sector, while the two branches split, remain separated by $d$, and recombine at distance $D$ from the black-hole center.}
\label{fig:dsw_setup}
\end{figure}

The Danielson--Satishchandran--Wald (DSW) thought experiment~\cite{Danielson:2022tdw,Danielson:2023yhu,Satishchandran:2025review}, shown in Figure~\ref{fig:dsw_setup}, places a supported probe in two spatially separated branches and later recombines them.  Their different tidal histories correlate with the unresolved black-hole sector and suppress exterior coherence after that sector is traced out.  The worldline noise kernel therefore determines the two-history decoherence functional.

Let the two physical histories be the SK branches $1,2$, with $\mathcal X^A_{a,L}=\mathcal X^A_{1,L}-\mathcal X^A_{2,L}$, $A\in\{\EE,\BB\}$.  Keeping this physical branch difference nonzero for the off-diagonal density matrix and collecting the active components into $\mathcal X_a$, the Gaussian influence functional gives $|\rho_{12}(t_f)/\rho_{12}(0)|=\exp[-\Gamma_{\rm dec}(t_f)]$, with
\begin{align}
  \Gamma_{\rm dec}(t_f)
  =&\frac12\sum_{A,B,L,L'}
  \int_{\tau_i}^{\tau_f}\dd\tau\int_{\tau_i}^{\tau_f}\dd\tau'\,
  \mathcal{X}^A_{a,L}(\tau)
  \NN^{AB}_{LL'}(\tau,\tau')
  \mathcal{X}^B_{a,L'}(\tau').
  \label{eq:Gamma_dec_general}
\end{align}
Here $\tau_i$ and $\tau_f$ are the worldline proper times bounding the protocol.\footnote{Exactly, $\tau_f-\tau_i=\int_0^{t_f}(\dd\tau/\dd t)\dd t$.  In weak-field PN coordinates, replacing this by $t_f$ is a 0PN approximation with relative $\ord(U/c^2,v^2/c^2)$ corrections; Eq.~\eqref{eq:Gamma_dec_general} itself retains proper time.}  The expression also applies to nonthermal and nonstationary Gaussian kernels, and positivity~\eqref{eq:N_positive} ensures $\Gamma_{\rm dec}\geq0$.

For stationary spectral estimates, in the leading weak-field rest frame $\tau_f-\tau_i\simeq t_f$, Eq.~\eqref{eq:Gamma_dec_general} becomes
\begin{equation}
  \Gamma_{\rm dec}(t_f)
  =\frac12\int\frac{\dd\omega}{2\pi}\,
  \mathcal{X}_{a}^{\dagger}(\omega)
  \NN(\omega)
  \mathcal{X}_{a}(\omega).
  \label{eq:Gamma_freq}
\end{equation}
For a split--hold--recombine history with approximately constant holding-stage difference, write
\begin{equation}
  \mathcal X_a(t)\simeq \mathcal X_{a,0}w_{t_f}(t),
  \qquad
  \Gamma_{\rm dec}(t_f)
  =\frac12\int\frac{\dd\omega}{2\pi}
  |\widetilde w_{t_f}(\omega)|^2
  \mathcal X_{a,0}^{\dagger}\NN(\omega)\mathcal X_{a,0},
  \label{eq:D_window_decoh}
\end{equation}
where $w_{t_f}$ is unity on the holding plateau and vanishes outside the protocol, with $\widetilde w_{t_f}(\omega)=\int\dd t\,\ee^{\ii\omega t}w_{t_f}(t)$.  Idealizing splitting and recombination as instantaneous gives the rectangular window $w_{t_f}=1$ for $0<t<t_f$ and zero otherwise,
\begin{equation}
  \widetilde w_{t_f}(\omega)
  =\ee^{\ii\omega t_f/2}
  \frac{2\sin(\omega t_f/2)}{\omega},
  \qquad
  \int\frac{\dd\omega}{2\pi}|\widetilde w_{t_f}(\omega)|^2=t_f.
  \label{eq:D_rectFT}
\end{equation}
If the projected noise is regular at the origin, keeping $\NN(\omega)=\NN(0)+\cdots$ over the selected frequencies gives
\begin{equation}
  \Gamma_{\rm dec}(t_f)
  \simeq \frac{t_f}{2}\,
  \mathcal X_{a,0}^{\dagger}\NN(0)\mathcal X_{a,0}.
  \label{eq:D_white}
\end{equation}
Thus the secular term is fixed by the zero-frequency noise; smooth switching changes only nonsecular contributions.

\subsubsection{Thermal Markovian regime and the DSW scaling}
\label{subsec:DSW}
For the DSW geometry the leading source is the electric quadrupolar tidal field.  Equal branch mass and fixed center of mass imply $I_{a,ij}=\mathcal O(md^2)$, hence at $D\gg d$, $\EE^{(0)}_{a,ij}=\mathcal O(Gmd^2/D^5)$ and $\EE_a^2=\mathcal O(G^2m^2d^4/D^{10})$; Appendix~\ref{app:decoherence} gives the coefficients.

For $r_h\ll D$, $m\ll M$ and slow switching, the secular term samples $|\omega|\lesssim t_f^{-1}\ll T_H$, where Eq.~\eqref{eq:common_schw_noise} is white.  Combining Eq.~\eqref{eq:D_white} with the DSW tidal scaling and suppressing the order-one geometry factor gives
\begin{equation}
  \Gamma_{\rm dec}^{\rm BH}(t_f)
  \sim
  \frac{G^6M^5m^2d^4}{\hbar\,c^{10}D^{10}}\,t_f
  \label{eq:DSW_scaling_units}
\end{equation}
In $G=\hbar=c=1$ units this is the DSW scaling $\Gamma_{\rm dec}^{\rm BH}\sim M^5m^2d^4t_f/D^{10}$~\cite{Danielson:2022tdw,Danielson:2023yhu,Danielson:2024yru}.  Appendix~\ref{app:decoherence} gives the geometry-dependent coefficient.  Related structures also arise in thermal-correlator~\cite{Biggs:2024doh}, horizon-radiation~\cite{Wilson-Gerow:2024ljx,Danielson:2024yru} and holographic~\cite{Kawamoto:2025decoherence} descriptions.

The soft interpretation follows from $\operatorname{Im}\DR\propto\omega$ and $\NN\simeq2\gamma T_H$: decoherence remains finite as $\omega\to0$, while absorbed energy carries extra powers of frequency, reproducing the soft-graviton picture~\cite{Danielson:2022tdw,Danielson:2023yhu}.  Beyond reproducing this rate, Eq.~\eqref{eq:D_white} shows that the interferometer isolates the zero-frequency symmetric correlator.  FDT ties it to tidal absorption only in equilibrium; away from equilibrium, response and noise are independent matching data.  Decoherence can therefore be compared with dissipative observables to test this relation, or extended to nonthermal and modified near-horizon sectors by rematching $\NN$.  The result also depends on how a finite-time protocol is completed~\cite{Danielson:2025minimize}.

\subsubsection{Beyond the white-noise approximation in the infrared}
\label{subsec:infrared_noise_decoh}

In a low-temperature quantum KMS regime, $\beta_q=T_q^{-1}$ and $\beta_q|\omega|\gg1$.  The Schwarzschild absorptive response remains Ohmic, $\operatorname{Im}\DR{}_{\EE}^{(2)}(\omega)\simeq\gamma_{\EE2}\omega$, so Eq.~\eqref{eq:quantum_noise_limit} gives the symmetric noise spectrum $\NN_{\EE}^{ij,kl}(\omega)\simeq\gamma_{\EE2}|\omega|\,\mathcal P_{\rm STF}^{ij,kl}$.\footnote{\label{fn:lowT_vacuum}Here $T_q$ is a generic KMS temperature and is not matched to the Schwarzschild Hartle--Hawking temperature $T_H$ in Eq.~\eqref{eq:hawkingT}.  The Boulware vacuum gives the $T_q\to0$ specialization discussed in Footnote~\ref{fn:vacuum_choice}.  Near-extremal charged or rotating black holes are another low-temperature setting we have in mind; extending the CTP matching to them is future work.}  Removing the common window factor in Eq.~\eqref{eq:D_window_decoh} leaves
\begin{equation}
  \mathcal E_{a,0}^{ij}\NN_{\EE\,ij,kl}(\omega)\mathcal E_{a,0}^{kl}
  \simeq \kappa_{\rm IR}|\omega|,
  \qquad
  \kappa_{\rm IR}\equiv
  \gamma_{\EE2}\mathcal E_{a,0}^{ij}\mathcal E_{a,0\,ij}.
  \label{eq:projected_ohmic_kernel}
\end{equation}

We use this linear-$|\omega|$ noise form only for $T_q\ll t_f^{-1}\lesssim|\omega|\le\omega_c$.  Here $\Lambda_{\rm EFT}$ is the EFT cutoff and $\tau_{\rm sw}$ is the characteristic switching time, so the controlled upper scale is chosen as $\omega_c\lesssim\min(\Lambda_{\rm EFT},\tau_{\rm sw}^{-1})$.  Substituting Eqs.~\eqref{eq:projected_ohmic_kernel} and \eqref{eq:D_rectFT} into Eq.~\eqref{eq:D_window_decoh} and integrating over $|\omega|\le\omega_c$ gives
\begin{equation}
  \begin{split}
  \Gamma_{\rm dec}(t_f)
  &=\frac{2\kappa_{\rm IR}}{\pi}
  \int_0^{\omega_c}\frac{\dd\omega}{\omega}
  \sin^2\!\left(\frac{\omega t_f}{2}\right)\\
  &=\frac{\kappa_{\rm IR}}{\pi}
  \left[\gamma_{\rm E}+\log(\omega_c t_f)-\operatorname{Ci}(\omega_c t_f)\right]\\
  &=\frac{\kappa_{\rm IR}}{\pi}\log(\omega_c t_f)+\ord(1),
  \qquad \omega_c t_f\gg1.
  \end{split}
  \label{eq:log_decoh}
\end{equation}
where $\gamma_{\rm E}$ is Euler's constant and $\operatorname{Ci}(x)\equiv-\int_x^\infty \dd t\,\cos t/t$ is the cosine integral.\footnote{The integral identity used in Eq.~\eqref{eq:log_decoh} is $\int_0^X \dd x\,(1-\cos x)/x=\gamma_{\rm E}+\log X-\operatorname{Ci}(X)$.}
Physically, Eq.~\eqref{eq:log_decoh} is the vacuum contribution of the soft horizon channel underlying the DSW soft-theorem/dressing picture~\cite{Danielson:2022tdw,Danielson:2023yhu}.  In the Boulware state there is no thermal population, but the branch-dependent tidal source can still excite arbitrarily soft black-hole modes that store which-path information, producing only logarithmic accumulation~\cite{Danielson:2024yru}.  In the Hartle--Hawking state the same modes are thermally populated, so Bose enhancement makes $\NN(\omega)$ white at $|\omega|\ll T_H$ and converts the logarithm into the linear-in-$t_f$ growth of Section~\ref{subsec:DSW}; temperature changes the infrared weighting of the same soft channel rather than the underlying mechanism.

\section{Outlook: microscopic quantum structure, matching across coarse grainings and gravitational splitting}
\label{sec:outlook}

The SK worldline EFT suggests extensions in three directions: microscopic information in the kernels, coarse-grained descriptions with spatial support beyond the worldline, and the definition of the gravitational subsystem itself.  We do not construct these matchings here.

\subsection{Microscopic quantum structure in worldline kernels}
\label{subsec:qg_matching}
\label{subsec:chaos_info}

Conventional dissipative worldline EFT matches the absorptive part of a retarded correlator to classical absorption~\cite{Goldberger:2005cd}.  The SK formulation enlarges the matching data to
\begin{equation}
  \operatorname{Re}\DR,
  \qquad \operatorname{Im}\DR,
  \qquad \NN,
  \qquad \mathcal K_{\SK}^{(n\ge3)},
  \label{eq:spectroscopy_hierarchy}
\end{equation}
which encode conservative response, dissipation, symmetric fluctuations and nonlinear/non-Gaussian information.  The present paper fixes only their low-frequency content; additional poles, thresholds or long-lived modes require separate matching and, when parametrically light, should be retained explicitly.

At the Gaussian two-point level, near-extremal black holes provide a concrete setting in which quantum near-horizon dynamics can modify the worldline response and noise kernels.  In near-AdS$_2$ JT/Schwarzian systems, low-energy correlators are dressed by the quantum density of states and transition matrix elements, so the low-frequency absorptive response need not remain simply Ohmic~\cite{MertensTuriaciVerlinde:2017schwarzian,Iliesiu:2020qvm,Brown:2024charged,EmparanTrezzi:2025transparency}.  Matching these quantum-corrected correlators onto $(\DR,\NN)$ would determine how microscopic near-horizon dynamics appears in absorption and decoherence~\cite{Biggs:2026notall}.

Beyond Gaussian matching, finer observables can access information that is invisible to two-point response and noise alone.  Here the finite microscopic state count becomes directly relevant: once level spacing or finer spectral correlations are resolved, smooth two-point kernels no longer contain all microscopic information.  Finite-entropy spectra, higher-point correlators and OTOCs probe discreteness, scrambling and spectral correlations; near-AdS$_2$/JT and its matrix-integral completions provide controlled settings for such effects~\cite{Maldacena:2015waa,Saad:2019lba,Stanford:2019vob,tHooft:1985,Terashima:2026brick}.  In SK language, this motivates extending the matching from $(\DR,\NN)$ to higher influence cumulants and other fine-grained spectral data.  Such information goes beyond reproducing Hawking thermodynamics or a single absorption coefficient~\cite{Biggs:2024doh,Biggs:2026notall}.

A separate direction is to use nonequilibrium kernels to characterize how a microscopic black-hole state approaches an effective stationary description.  During formation, appreciable absorption or evaporation, their central-time dependence tracks relaxation; the subsequent emergence of stationarity and the KMS condition would indicate coarse-grained equilibration of the observables retained by the EFT~\cite{KourkoulouMaldacena:2017pure,BhattacharyaJatkarSorokhaibam:2019quench}.  This diagnoses only coarse-grained equilibration of the retained observables and does not require the full microscopic state to become Gibbs.  Appendix~\ref{app:noneq} gives the corresponding slow-evolution expansion.

\subsection{Matching to spatially extended coarse-grained descriptions}
\label{subsec:coarse_graining}

The distinction here is not whether finite-size effects are retained.  The worldline EFT already incorporates them systematically through multipole operators, Wilson coefficients and a derivative expansion.  Rather, after the strong-gravity region is integrated out, it assumes that the coarse-grained degrees of freedom and their couplings are supported only on the worldline, so spatial separation inside the compact object no longer appears as an independent argument of the effective correlators.  Relaxing this assumption allows the coarse-grained variables and interactions themselves to have higher-dimensional spatial support, retaining gradients and correlations between distinct spatial points.

Membrane and fluid descriptions provide an intermediate example.  Their effective variables and interactions live on a stretched horizon or finite-cutoff hypersurface, so spatial gradients and correlations along that surface remain explicit.  The Damour--Navier--Stokes description organizes classical horizon transport in this form~\cite{Price:1986yy,Thorne:1986membrane,Gourgoulhon:2005dns}, while Wilsonian fluid/gravity shows that long-wavelength gravitational perturbations at a finite radial cutoff obey hydrodynamic dispersion on the cutoff surface, approaching the membrane description toward the horizon~\cite{Bredberg:2011wilsonian}.  A stochastic or SK completion can additionally retain fluctuating stresses and correlators between distinct points on the horizon~\cite{Crossley:2015evo,Glorioso:2017fpd,Haehl:2015pja}.  More general near-horizon descriptions need not use membrane variables and can instead specify frequency-dependent boundary response, including partial reflectivity~\cite{Datta:2019euh,Oshita:2019reflectivity}.  Such descriptions retain spatial and propagation information that has already been integrated out in the worldline theory; sufficiently persistent reflected components can, for example, produce late-time echo structure~\cite{Cardoso:2016echo,Oshita:2019reflectivity}.

A concrete illustration is dissipative dispersion.  Ref.~\cite{Oshita:2019reflectivity} introduces a spatial-gradient dissipative term in the near-horizon wave equation, motivated by viscous sound propagation, while Ref.~\cite{LiberatiMaccione:2014} gives a related hydrodynamic model with an imaginary cubic-in-momentum correction to the low-energy dispersion relation.  These are local partial differential equation (PDE) models rather than spatially nonlocal kernels in the strict sense, but they show how additional gradient structure can remain explicit once the effective variables and interactions are not restricted to worldline support.

A bulk description retains spatial dependence throughout the curved spacetime.  In stochastic gravity, the Einstein--Langevin equation keeps spacetime-dependent stress-tensor fluctuations, stochastic metric perturbations and their propagation on the full background~\cite{Calzetta:1986ctp,Hu:2008rga,Sinha:2002stoch,HuRoura:2007}.  The resolved metric also propagates in the coupled classical dynamics of Section~\ref{subsec:recoil} and Appendix~\ref{app:hs_langevin}, but the black-hole-sector noise source remains supported on the worldline; a bulk noise kernel retains correlations between distinct bulk points before compact-object reduction.  In their overlapping long-wavelength regime, horizon or bulk correlators can be projected onto compact-object multipole channels, while the corresponding worldline Wilson data encode the result after the additional spatial structure has been integrated out.  Matching at common scales is therefore the link among descriptions with different spatial support.

\subsection{Gravitational splitting and subsystem definition}
\label{subsec:qrf_edge}

Any influence-functional description in gravity presupposes a \emph{gravitational splitting}: one must specify which observables or algebras are retained and which gravitational degrees of freedom are traced over.  Unlike a fixed-background tensor-factor decomposition, this split is constrained by gravitational dressing and gauge constraints~\cite{Donnelly:2016auv,Speranza:2017gxd,Freidel:2020svx,Donnelly:2017jcd}.  Edge-mode and dynamical-frame constructions refine this notion by supplementing the subregion split with boundary or reference-frame data~\cite{Carrozza:2021gju,CarrozzaEcclesHoehn:2022frames,GoellerHoehnKirklin:2022frames,FreidelKirklin:2025localization}.  Once a split is specified, the SK kernels $(\DR,\NN)$ describe the corresponding reduced subsystem, so response, noise and reduced-state decoherence are relative to that choice rather than statements of fundamental nonunitarity.

Quantum-reference-frame (QRF) theory sharpens this subsystem dependence.  Different internal QRFs can induce different subsystem algebras and tensor-product structures~\cite{Ahmad:2021qrfsubsystems}, and even whether a subsystem appears open or closed can be QRF relative~\cite{HoehnKotechaMele:2023qrf}.  Ref.~\cite{Cepollaro:2024tradeoff} shows, in ideal QRF models, that coherence can be converted into entanglement under a frame change while operational probabilities remain invariant.  This provides a useful template for gravity: an apparent loss of coherence assigned to one subsystem need not represent invariant information loss.  In superposed semiclassical geometries, localization and branch identification are likewise QRF dependent~\cite{deLaHamette:2021qrf,KabelEtAl:2024identification}, while perturbative-gravity constructions realize the corresponding relational observables through boundary or matter reference fields~\cite{KabelBruknerWieland:2023boundary,GoellerHoehnKirklin:2022frames,ChenGiacomini:2026qrfgravity}.

The DSW setup makes the issue operational.  Its branch-dependent horizon/soft record carries which-path information~\cite{Danielson:2025softQI}, but saying that this information has ``left the probe'' already assumes a split among the probe, resolved gravitational field and horizon/soft sector.  Field-theoretic treatments of quantum split sources make the associated superposition of gravitational configurations explicit~\cite{ChenGiacominiRovelli:2022split}, while QRF analyses of gravity-mediated entanglement show that the identification of the mediating gravitational degree of freedom is relational~\cite{ChristodoulouEtAl:2022gravityent}.  More generally, gauge constraints need not eliminate operationally meaningful local algebras even when no spacetime-local tensor-product factorization exists~\cite{FewsterEtAl:2024qrfqft,SpalvieriEtAl:2025localops}.  A QRF-covariant SK description should therefore transform the source operators, accessible algebra and reduction map together while preserving relational interference probabilities.

Two extensions then follow directly.  First, one can ask in perturbative gravity how the coherence--entanglement trade-off changes under a QRF or gravitational-split transformation, tracking the redistribution of correlations among the probe, resolved gravitons and horizon/soft sector while keeping the relational interference visibility fixed~\cite{Cepollaro:2024tradeoff,KabelBruknerWieland:2023boundary,GoellerHoehnKirklin:2022frames,ChenGiacomini:2026qrfgravity}.  Second, for a fully dynamical interference protocol one should determine which relational observables map onto the worldline kernels: at Gaussian order this identifies the multipole and frequency projection of $\NN$ actually measured by the protocol, while beyond Gaussian order the same question extends to higher SK cumulants.  Matching these effective observables to microscopic horizon or soft-graviton correlators would specify what black-hole quantum information is, in principle, accessible to DSW-type experiments.

The subsystem choice can thus be regarded as part of the matching problem.  A distinct issue arises in evaporation, where island/quantum extremal surface (QES) constructions show that the bulk region reconstructible from a radiation subsystem can change in time~\cite{Almheiri:2019psf,Penington:2019npb}; relating such global reconstruction data to the local SK influence functional remains open.

\section{Conclusions}
\label{sec:conclusions}

We have formulated a Schwinger--Keldysh (SK) worldline EFT in which unresolved black-hole multipoles couple to exterior tidal fields~\eqref{eq:tidal_coupling} and are integrated out at the level of the closed-time-path generating functional.  In the spherical reference sector their STF one-point functions vanish~\eqref{eq:zero_mean}, while their connected two-point functions generate the quadratic influence functional~\eqref{eq:quadratic_IF}.  This identifies the retarded kernel $\DR$ and symmetric kernel $\NN$ as the two real-time Wilson data controlling, respectively, causal tidal response and fluctuations/decoherence.  Their common microscopic origin does not imply that one determines the other out of equilibrium.  Instead, for stationary states the spectral representation and Wightman positivity give the nonthermal bounds~\eqref{eq:positivity_bound} and~\eqref{eq:scalar_bound}, while the Hubbard--Stratonovich representation turns the positive kernel $\NN$ into a stochastic source whose effects propagate through the resolved worldline and metric.  In this way conservative/absorptive response and stochastic fluctuations are organized within the same worldline EFT rather than introduced as separate phenomenological sectors.

For a Gibbs reference state, microscopic KMS yields the quantum fluctuation--dissipation theorem (FDT)~\eqref{eq:FDT}; for equilibrium Schwarzschild matching the temperature is $T_H$~\eqref{eq:hawkingT}.  In the local low-frequency regime, the same equilibrium input is encoded by dynamical KMS and yields the Einstein relation~\eqref{eq:Einstein_local}, while the full frequency-space FDT remains the more general statement.  These equilibrium relations constrain the reference correlators and need not remain valid after the system is driven.  A stationary nonthermal state instead has independently matched response and noise, constrained only by positivity, while genuinely evolving states require two-time kernels.  When the evolution is slow, the Wigner expansion~\eqref{eq:wigner} organizes the gradient corrections, and a local FDT follows only with the additional assumption of local KMS.  The framework therefore cleanly separates stationarity, thermality and adiabatic local equilibrium, and provides a fluctuation--response diagnostic of equilibration rather than imposing thermal behavior by construction.

A central application is obtained by propagating the Hubbard--Stratonovich source through the resolved metric and worldline as one coupled random system.  Direct and metric-mediated forces are coherent pieces of the same realization and generate nonzero metric--orbit cross correlations.  At the first nonvanishing post-Newtonian order of the Schwarzschild electric-quadrupole noise sector, these terms exhaust the relative-acceleration source, restore center-of-mass momentum balance realization by realization, and give Eq.~\eqref{eq:main_feedback_full}.  Restoring $\hbar$ gives the explicit $\ell_{\rm P}^2$ suppression in Eq.~\eqref{eq:main_feedback_planck}, while the accumulated phase fluctuation remains Planck suppressed within the controlled inspiral window.  This is a complete leading-PN source-level result rather than an all-orders evaluation of $C_{zz}$; turning it into a fully propagated stochastic orbital observable requires a dressed nonstationary propagator, coincident self-field renormalization, higher-PN potential terms, radiation/tails and the full frequency-dependent noise kernel.

The DSW application gives the symmetric kernel $\NN$ a direct reduced-state interpretation.  For a supported spatial superposition, the two branches generate different tidal histories, and tracing over the unresolved black-hole sector contracts their difference directly with $\NN$ through~\eqref{eq:Gamma_dec_general}.  Schwarzschild absorption matching together with equilibrium FDT then reproduces the known horizon-induced scaling~\eqref{eq:DSW_scaling_units}~\cite{Danielson:2022tdw,Biggs:2024doh}.  At Gaussian order, the worldline formulation isolates the quantity entering the decoherence functional: a protocol-dependent projection of the symmetric multipole noise kernel $\NN$, rather than the retarded response alone.  Away from equilibrium, where response and noise need not satisfy FDT, DSW-type interference is therefore, in principle, sensitive to states with the same absorptive response but different noise spectra.  Tidal heating and stochastic orbital backreaction on the one hand, and DSW decoherence on the other, are thus complementary probes of the same matched black-hole data rather than separate phenomenological noise models.  A further direction is to solve a quantum master equation for the reduced state of the probe superposition or exterior gravitons and ask how much of the black-hole influence functional can be inferred from their time evolution.

Several extensions broaden the framework toward quantum black-hole physics.  Near-extremal and finite-entropy models can determine quantum corrections to the two-point kernels and motivate matching to higher SK data and nonequilibrium evolution~\cite{MertensTuriaciVerlinde:2017schwarzian,Iliesiu:2020qvm,EmparanTrezzi:2025transparency}.  A second direction matches the worldline theory to spatially extended coarse-grained descriptions: stretched-horizon/fluid descriptions~\cite{Thorne:1986membrane,Bredberg:2011wilsonian} and bulk Einstein--Langevin descriptions~\cite{Calzetta:1986ctp,Hu:2008rga,Sinha:2002stoch,HuRoura:2007} retain gradients and correlations that are integrated out in the worldline reduction.
Gravitational splitting raises the question of how reduced gravitational subsystems and observables should be defined when gauge constraints, dressing and quantum reference frames make their identification relational~\cite{Donnelly:2016auv,Speranza:2017gxd,Ahmad:2021qrfsubsystems,HoehnKotechaMele:2023qrf,Cepollaro:2024tradeoff}.
This issue is especially relevant for black-hole quantum dynamics, as illustrated by DSW-type interference and by island/QES constructions in evaporating black holes~\cite{KabelBruknerWieland:2023boundary,ChenGiacomini:2026qrfgravity,Almheiri:2019psf,Penington:2019npb}.
The worldline influence functional provides a concrete low-energy starting point for formulating such relational questions in terms of explicit dynamical protocols and observables.

Taken together, the astrophysical response sector fixes and tests the low-energy kernels, while stochastic backreaction and DSW-type decoherence expose their fluctuation and reduced-state content.  The worldline SK EFT therefore provides a common low-energy language for connecting causal response, noise and decoherence to microscopic black-hole dynamics, while keeping the state dependence of the corresponding kernels explicit.  Extending this matching beyond Gaussian equilibrium data should make it possible to ask systematically which aspects of black-hole quantum dynamics can be inferred from exterior measurements without committing to a particular microscopic model.

\section*{Acknowledgements}
\label{sec:Acknowledgements}

The author thanks Shinji Mukohyama and Kanji Nishii for useful comments and discussions.
The author also thanks Toshifumi Noumi, Masataka Ishikawa, Kanji Nishii, and Kaho Yoshimura for helpful correspondence regarding this manuscript and their related work.
This work was supported by JST  (Japan Science and Technology Agency) SPRING, Grant No.\ JPMJSP2110.

During the preparation of this manuscript, the author used ChatGPT (OpenAI) to assist
with language editing, organization of the text, and checking the clarity of explanations. The
author reviewed and edited all AI-assisted output and takes full responsibility for the content
of the manuscript.
\appendix
\section{Quadratic influence functional and conventions}
\label{app:influence}

For a single Hermitian internal operator $Q$ coupled to a c-number tidal source $\mathcal X$,
\begin{equation}
  S_{\rm int}=\int\dd\tau\,Q(\tau)\mathcal X(\tau),
  \label{eq:A_Sint}
\end{equation}
the SK influence functional is
\begin{equation}
  Z_{\IF}[\mathcal X_1,\mathcal X_2]
  =\avg{T_C\exp\!\left[\ii\int_C\dd\tau\,Q(\tau)\mathcal X(\tau)\right]}_{\rho_{\rm BH}^{\rm init}}
  \equiv\ee^{\ii S_{\IF}},
  \label{eq:A_ZIF}
\end{equation}
where the contour integral gives $+\int\mathcal X_1Q_1$ on the forward branch and $-\int\mathcal X_2Q_2$ on the backward branch.  Define $S^>(\tau,\tau')\equiv\avg{Q(\tau)Q(\tau')}_{\rho_{\rm BH}^{\rm init}}$ and $S^<(\tau,\tau')\equiv\avg{Q(\tau')Q(\tau)}_{\rho_{\rm BH}^{\rm init}}$.  For $\avg{Q}_{\rho_{\rm BH}^{\rm init}}=0$, the connected quadratic cumulant is
\begin{equation}
  \ii S_{\IF}^{(2)}
  =-\frac12\sum_{a,b=1}^2\eta_a\eta_b
  \int\dd\tau\dd\tau'\,
  \mathcal X_a(\tau)G^{ab}(\tau,\tau')\mathcal X_b(\tau'),
  \qquad
  \eta_1=+1,\quad\eta_2=-1,
  \label{eq:A_cumulant}
\end{equation}
with
\begin{align}
 G^{11}&=\theta(\tau-\tau')S^>+\theta(\tau'-\tau)S^<,
 &G^{22}&=\theta(\tau'-\tau)S^>+\theta(\tau-\tau')S^<,\\
 G^{12}&=S^<,
 &G^{21}&=S^>,
 \label{eq:A_SKmatrix}
\end{align}
where the arguments $(\tau,\tau')$ of $S^{>,<}$ are implicit.

Using
\begin{equation}
  \mathcal X_r=\frac{\mathcal X_1+\mathcal X_2}{2},
  \qquad
  \mathcal X_a=\mathcal X_1-\mathcal X_2,
  \label{eq:A_ra}
\end{equation}
and $S^>-S^<=\avg{[Q,Q]}_{\rho_{\rm BH}^{\rm init}}$, $S^>+S^<=\avg{\{Q,Q\}}_{\rho_{\rm BH}^{\rm init}}$, one obtains
\begin{equation}
  S_{\IF}^{(2)}
  =\int\dd\tau\dd\tau'\,\mathcal X_a(\tau)\DR(\tau,\tau')\mathcal X_r(\tau')
  +\frac{\ii}{2}\int\dd\tau\dd\tau'\,\mathcal X_a(\tau)\NN(\tau,\tau')\mathcal X_a(\tau'),
  \label{eq:A_IF_final}
\end{equation}
with
\begin{align}
  \DR(\tau,\tau')&=\ii\theta(\tau-\tau')\avg{[Q(\tau),Q(\tau')]}_{\rho_{\rm BH}^{\rm init}},\\
  \NN(\tau,\tau')&=\frac12\avg{\{Q(\tau),Q(\tau')\}}_{\rho_{\rm BH}^{\rm init}}.
  \label{eq:A_kernels}
\end{align}
This uses the same $+\ii\theta[Q,Q]$ convention as the main text.  If $\avg{Q}_{\rho_{\rm BH}^{\rm init}}\neq0$, the one-point function adds a term linear in $\mathcal X_a$ and $\NN$ is defined from the connected covariance.

Restoring $Q\to Q_A^L$ and $\mathcal X\to\mathcal X_L^A$, with the channel and multipole sums, reproduces Eq.~\eqref{eq:quadratic_IF}; the conventional multipolar combinatorial factors are already absorbed into $Q_A^L$.  Rotational invariance and parity then give the channel structure described in Section~\ref{sec:sk}; spin or parity violation can introduce additional mixing and requires separate matching.

\section{Spectral representation, positivity and KMS}
\label{app:spectral}

For the reference correlators defined in Section~\ref{sec:worldline}, stationarity means
$[\rho_{\rm BH}^{\rm init},H_{\rm BH}]=0$.  For such a stationary reference state, the Fourier transforms of the Wightman functions in Eq.~\eqref{eq:wightman_fourier} are positive matrix-valued spectral distributions.  Equivalently, the quadratic form obtained by smearing either ordering with arbitrary test functions is nonnegative.  If the spectrum is absolutely continuous this becomes a positive-semidefinite density almost everywhere, whereas a pure-point component appears as a sum of delta functions.  Thus, without assuming spectral discreteness,
\begin{equation}
  \rho_Q=S^>-S^<=2\,\operatorname{Im}\DR,
  \qquad
  \NN=\frac{S^>+S^<}{2},
  \qquad
  \NN\pm\operatorname{Im}\DR=S^{>,<}\succeq0,
  \label{eq:B_positivity}
\end{equation}
where the last relation is understood at the level of spectral measures and proves Eq.~\eqref{eq:positivity_bound}.

For the ETH discussion in Section~\ref{subsec:eth}, it is useful to display the pure-point specialization explicitly.  If
\begin{equation}
  H_{\rm BH}|n\rangle=E_n|n\rangle,
  \qquad
  \rho_{\rm BH}^{\rm init}=\sum_n p_n|n\rangle\langle n|,
  \label{eq:B_discrete_basis}
\end{equation}
then, for one Hermitian multipole operator $Q$,
\begin{align}
 S^>(\omega)
 &=2\pi\sum_{m,n}p_n|Q_{mn}|^2
 \delta\!\left(\omega-E_m+E_n\right),
 \\
 S^<(\omega)
 &=2\pi\sum_{m,n}p_m|Q_{mn}|^2
 \delta\!\left(\omega-E_m+E_n\right).
 \label{eq:B_spectral}
\end{align}
Continuous or mixed spectra are represented instead by the corresponding spectral integrals; Eqs.~\eqref{eq:B_positivity} and the KMS relations below are unchanged.

For a Gibbs reference state, $\rho_{\rm BH}^{\rm init}=Z^{-1}e^{-\beta H_{\rm BH}}$, the time-domain KMS condition gives, independently of whether the spectrum is discrete or continuous,
\begin{equation}
  S^<(\omega)=e^{-\beta\omega}S^>(\omega).
  \label{eq:B_detailed}
\end{equation}
In the pure-point form above this is transparent term by term, since the delta function fixes $E_m-E_n=\omega$ and hence $p_m=e^{-\beta\omega}p_n$.  Combining Eq.~\eqref{eq:B_detailed} with Eq.~\eqref{eq:B_positivity} gives
\begin{align}
  S^>(\omega)&=\frac{2}{1-e^{-\beta\omega}}
  \operatorname{Im}\DR(\omega),&
  S^<(\omega)&=\frac{2}{e^{\beta\omega}-1}
  \operatorname{Im}\DR(\omega),\\
  \NN(\omega)&=\coth\!\left(\frac{\beta\omega}{2}\right)
  \operatorname{Im}\DR(\omega).
  \label{eq:B_FDT}
\end{align}
These equations are distributional when delta-function contributions are present; the Wightman spectral measures remain nonnegative.

For a spinning and/or charged extension, suppose a suitable mode sector admits a grand-canonical KMS state,
\begin{equation}
  \rho\propto
  \exp\!\left[-\beta_H\left(H-\Omega_HJ-\Phi_H Q_{\rm el}\right)\right].
  \label{eq:B_grandcanonical}
\end{equation}
A mode with quantum numbers $(\omega,m,q)$ then obeys detailed balance in terms of the horizon-frame frequency
\begin{equation}
  \widetilde\omega
  =\omega-m\Omega_H-q\Phi_H,
  \label{eq:B_tildeomega}
\end{equation}
so that, under this KMS assumption, the thermal factor becomes $\coth(\beta_H\widetilde\omega/2)$.  This formal density matrix does not assert the existence of a globally regular Hartle--Hawking state for asymptotically flat Kerr~\cite{Kay:1988mu}.  Superradiant channels have $\widetilde\omega<0$; their apparent negative absorption is consistent with Wightman positivity because the sign of the commutator spectral distribution changes together with the thermal factor.  Applying these mode relations to a rotating hole requires the spinning worldline CTP and independent matching of its response and noise, beyond the spherical construction in Section~\ref{sec:worldline}~\cite{Goldberger:2020yvj}.

\section{Two-time kernels and Wigner expansion away from equilibrium}
\label{app:noneq}

This appendix provides the technical implementation of the nonequilibrium discussion in Section~\ref{subsec:thermalization}.  In particular, it makes precise when the two-time kernels introduced there acquire central-time dependence, how the slow-evolution limit is organized, and what can be inferred from a local fluctuation--response ratio.  In general $F(\tau,\tau')=\Tr_{\rm BH}[\rho_{\rm BH}^{\rm init}\widehat F(\tau,\tau')]$.  If $[\rho_{\rm BH}^{\rm init},H_{\rm BH}]=0$ and the source-free black-hole Hamiltonian $H_{\rm BH}$ is proper-time independent, the kernels are stationary even for a nonthermal state; only the Gibbs specialization implies KMS.  Central-time dependence instead arises from a nonstationary initial state or from self-consistent evolution of the reference background/state.  A weak source treated in linear response around a fixed thermal reference state does not by itself require time-dependent kernels.

For slow evolution define
\begin{equation}
  \bar\tau=\frac{\tau+\tau'}{2},\qquad s=\tau-\tau'
  \label{eq:E_Ts}
\end{equation}
and use the Wigner transform~\eqref{eq:wigner}.  The convolution
\begin{equation}
  (A\circ B)(\tau,\tau')=\int\dd\tau_1\,A(\tau,\tau_1)B(\tau_1,\tau')
  \label{eq:E_convolution}
\end{equation}
becomes
\begin{equation}
  (A\circ B)(\bar\tau,\omega)
  =A\exp\!\left[\frac{\ii}{2}
  \left(
  \overleftarrow{\partial_\omega}\overrightarrow{\partial_{\bar\tau}}
  -\overleftarrow{\partial_{\bar\tau}}\overrightarrow{\partial_\omega}
  \right)\right]B .
  \label{eq:E_Moyal}
\end{equation}
The sign of the Moyal exponent follows from the $e^{+\ii\omega s}$ Fourier convention in~\eqref{eq:wigner}.  If the macroscopic evolution time $\tau_{\rm rel}$ is much longer than the correlation time $\tau_c$, gradients are suppressed by $\tau_c/\tau_{\rm rel}$ and the zeroth-order theory is locally stationary.  This appendix assumes that hierarchy; it does not provide a controlled expansion for a kernel with a parametrically long correlation time comparable to the resolved evolution time.

When local equilibrium is valid,
\begin{equation}
  \NN(\bar\tau,\omega)
  =\coth\!\left(\frac{\beta(\bar\tau)\omega}{2}\right)
  \operatorname{Im}\DR(\bar\tau,\omega)
  +\ord(\partial_{\bar\tau}).
  \label{eq:E_localFDT}
\end{equation}
A slowly varying surface gravity alone is insufficient; the probed operator sector must also remain close to local equilibrium.

For a generic nonequilibrium state define
\begin{equation}
  R(\bar\tau,\omega)
  \equiv
  \frac{\NN(\bar\tau,\omega)}{\operatorname{Im}\DR(\bar\tau,\omega)},
  \label{eq:E_R}
\end{equation}
and, for $\omega>0$ in a diagonal bosonic channel where $\operatorname{Im}\DR>0$ and $R>1$,
\begin{equation}
  \beta_{\rm eff}(\bar\tau,\omega)
  =\frac{2}{\omega}\operatorname{arccoth}R(\bar\tau,\omega).
  \label{eq:E_betaeff}
\end{equation}
The definition applies only on this frequency range; for negative frequencies the ratio must be treated with the corresponding sign convention.  Even when the ratio can be inverted, $\beta_{\rm eff}^{-1}$ is a thermodynamic temperature only if it is approximately independent of frequency and operator and the other local-equilibrium relations hold.

The nonstationary decoherence functional remains
\begin{equation}
  \Gamma_{\rm dec}
  =\frac12\int\dd\tau\dd\tau'\,
  \Delta\mathcal X(\tau)\NN(\tau,\tau')\Delta\mathcal X(\tau'),
  \label{eq:E_noneqdecoh}
\end{equation}
and the covariance of the HS auxiliary field is likewise the exact $\NN(\tau,\tau')$.  These expressions are the formal two-time influence-functional relations for nonstationary microscopic states.  Within the controlled slow-evolution regime of this appendix, they can be used to test whether central-time dependence decreases and whether the probed operator sector approaches local KMS.  Determining a generic long-time memory tail would instead require the full low-frequency spectral structure beyond the gradient data retained here.

\section{Coupled stochastic metric--worldline dynamics}
\label{app:hs}
\label{app:hs_langevin}

\subsection{Covariant sources and coupled retarded solution}
\label{app:covariant_coupled_solution}

Equation~\eqref{eq:HS} introduces an auxiliary classical Gaussian field $\xi_A^L$ whose mean vanishes and whose covariance is~\eqref{eq:xi_cov}.  The indices $A,L$ identify the same tidal channel and symmetric-trace-free tensor structure as the operator $Q_A^L$; $\xi_A^L$ is not a new black-hole degree of freedom.  To vary the worldline and metric covariantly, introduce a common parameter $\lambda$ and
\begin{equation}
 q_{(b)}=\sqrt{-g^{(b)}_{\mu\nu}(z_b)\dot z_b^\mu\dot z_b^\nu},\qquad
 q_r=\frac{q_{(1)}+q_{(2)}}2,\quad q_a=q_{(1)}-q_{(2)},\quad \dd\tau=q_r\dd\lambda .
 \label{eq:C_branch_measure}
\end{equation}
To first order in difference fields the stochastic action is
\begin{equation}
 S_\xi^{(1)}=\sum_{A,L}\int\dd\lambda\,\xi_A^L
 \left(q_r\mathcal X^A_{a,L}+q_a\mathcal X^A_L\right).
 \label{eq:C_covariant_reduced_action}
\end{equation}
The physical stochastic equations are obtained by varying with respect to the difference fields $h_a$ and $z_a$ while holding the $r$-history and the realization $\xi_A^L$ fixed, and only then imposing the physical limit $h_a=z_a=0$.  In the worldline variation, $z_a$ enters both the tidal pullback $\mathcal X_{a,L}^A$ and the measure $q_a$.  Equivalently, varying the two branch actions before taking the physical limit gives the ordinary covariant Euler--Lagrange derivative of $qL_\xi$ evaluated on the $r$-history.  Thus
\begin{equation}
 \mathscr J_\xi^{\mu\nu}(x)=\left.\frac{\delta S_\xi}{\delta h_{a\mu\nu}(x)}\right|_{a=0},\qquad
 F_{\xi\mu}=\frac1q\left[\frac{\partial(qL_\xi)}{\partial z^\mu}
 -\frac{\dd}{\dd\lambda}\frac{\partial(qL_\xi)}{\partial\dot z^\mu}\right]_r,
 \qquad L_\xi=\sum_{A,L}\xi_A^L\mathcal X_L^A ,
 \label{eq:C_noise_sources}
\end{equation}
where $q$ reduces to $q_r$ in the physical limit.  The derivative in $F_{\xi\mu}$ is the full worldline variation: it includes the proper-time measure, the velocity dependence of the tidal tensors and the variation of the comoving frame.  If only the explicit position dependence of the tidal pullback is varied at fixed $u^\mu$ and $e_i^\mu$, one obtains the simpler piece $F_\xi^\mu|_{u,e}=\sum_{A,L}\xi_A^L\nabla^\mu\mathcal X_L^A$.  Similarly, varying $h_a$ acts on both the tidal pullback and $q_a$, so the metric source contains both contributions,
\begin{equation}
 \mathscr J_\xi^{\mu\nu}(x)=\sum_{A,L}\int\dd\lambda\,\xi_A^L
 \left[q_r\frac{\delta\mathcal X^A_{a,L}}{\delta h_{a\mu\nu}(x)}
 +\mathcal X^A_L\frac{\delta q_a}{\delta h_{a\mu\nu}(x)}\right]_{a=0}.
 \label{eq:C_metric_noise_source}
\end{equation}
The shorthand $Q^L_{A,\rm eff}=\avg{Q_A^L}_{\mathcal X_r}+\xi_A^L$ is therefore useful only at the source level: the metric produced by $\xi$ must still be propagated and reinserted into the worldline equation.

Before turning on $\xi$, the deterministic SK equations give the retarded in-in metric one-point function $\langle h_{\mu\nu}\rangle$, obtainable by Green-function iteration or classical tree diagrams~\cite{Goldberger:2004jt,IvanovZhou:2023matching}.  Matching it to black-hole perturbations fixes the response, including the nonzero-frequency dynamical sector~\cite{CombaluzierSzteinsznaider:2025dyn,Kobayashi:2025vgl}.

Let $(\bar h,\bar z)$ solve these deterministic equations at $\xi=0$.  We choose de Donder gauge, $\bar\nabla^\mu(h_{\mu\nu}-\tfrac12\bar g_{\mu\nu}h)=0$, to fix the retarded propagator and compare with the nonzero-frequency EFT one-point-function matchings of Refs.~\cite{CombaluzierSzteinsznaider:2025dyn,Kobayashi:2025vgl}, which use the same gauge.  Equation~\eqref{eq:linear_coupled_noise_blocks} then holds with $L_{hh}=L_{hh}^{\rm dD}$ and
\begin{equation}
 G_h^R\equiv(L_{hh}^{\rm dD})^{-1}_{\rm ret},\qquad
 h_\xi=G_h^R(B_h\xi-L_{hz}z_\xi).
 \label{eq:C_metric_solution}
\end{equation}
The Schur-complement solution is exactly~\eqref{eq:noise_transfer_main}.  The random tidal field $\delta_h\mathcal X[h_\xi]$ and the induced variation $\delta\avg{Q_A^L}(\tau)=\int\dd\tau'\,\DR{}^{AB}_{LL'}(\tau,\tau')\,\delta\mathcal X_B^{L'}(\tau')$ belong to the deterministic $L$ blocks; adding them again as a new random force would double count the same realization.  Gauge consistency is likewise inherited from the common diffeomorphism-invariant action: before gauge fixing the metric and worldline variations obey the Ward identity associated with the Bianchi identity.

After the fixed-realization solution is known, the stationary worldline spectrum is
\begin{equation}
 C_{zz}(\omega)=G_{z,\rm eff}^R B_{\rm eff}\,\NN(\omega)\,
 B_{\rm eff}^\dagger G_{z,\rm eff}^A,
 \qquad B_{\rm eff}=B_z-L_{zh}G_h^RB_h,
 \label{eq:C_worldline_spectrum}
\end{equation}
where $G_{z,\rm eff}^R\equiv(L_{\rm eff})^{-1}_{\rm ret}$ is the effective worldline retarded propagator introduced in Eq.~\eqref{eq:noise_transfer_main}, $G_{z,\rm eff}^A\equiv(G_{z,\rm eff}^R)^\dagger$, and all frequency arguments are understood.  The fixed-metric reduction instead sets the stochastic metric response to zero and uses the undressed worldline propagator $G_z^R\equiv(L_{zz})^{-1}_{\rm ret}$:
\begin{equation}
 C_{zz}^{\rm fix}=G_z^R B_z\NN B_z^\dagger G_z^A,
 \qquad G_z^A\equiv(G_z^R)^\dagger .
 \label{eq:C_fixed_metric_reduction}
\end{equation}
Thus the complete Gaussian-linear transfer solution replaces both the response operator and the forcing map; it is not the fixed-metric covariance plus an independent ``metric noise'' term.  Even if $\NN(\omega)\simeq2T_H\gamma$ is white at source level, the transfer functions generally make $h_\xi$ and $z_\xi$ colored.

The coincident part of $G_h^RB_h\xi$ evaluated on the same worldline is singular.  A complete treatment requires regulating this stochastic self-field and matching the resulting local terms onto renormalized worldline coefficients, analogously to deterministic self-force EFT~\cite{GalleyHu:2009}.  We do not perform that coincident-field renormalization here; the leading-PN result below keeps only the finite inter-body propagation.

\subsection{Leading-PN Schwarzschild binary and semiclassical scaling}
\label{app:pn_coupled_noise}

We now specialize the general coupled system to the leading-order binary example used in Section~\ref{subsec:recoil}.  In this subsection $I,J\in\{1,2\}$ label the two bodies; because only the electric $\ell=2$ channel is retained, the tidal-channel label used above is suppressed.  Each body is a nonspinning Schwarzschild black hole, and the response is expanded in the long-wavelength regime $GM_I|\omega|\ll1$.  The phrase ``leading PN'' below refers to the first nonvanishing term within this horizon-noise finite-size sector.  With $v^2\sim GM_{\rm tot}/r$ and $\Omega r\sim v$, the standard potential-region counting~\cite{Goldberger:2004jt,Porto:2016pyg} gives
\begin{equation}
 \frac{k^0}{|\bm k|}\sim v,\qquad
 G_{h,\rm pot}^R(k^0,\bm k)=G_{h,\rm pot}^{(0)}(\bm k)\left[1+\ord(v^2)\right],\qquad
 \mathcal E_{ij}\sim\frac{Gm}{r^3},\quad \mathcal B_{ij}\sim v\mathcal E_{ij}.
 \label{eq:C_PN_power_counting}
\end{equation}
Thus the instantaneous electric quadrupole gives the leading finite inter-body stochastic interaction.  Retardation and 1PN electric corrections are relative $\ord(v^2)$, magnetic noise enters the covariance at relative $\ord(v^2)$, and radiation/tails are further suppressed.  Also $\NN(\omega)=\NN(0)[1+\ord((GM_I\omega)^2)]$ with $GM_I\Omega\sim v^3$ for comparable masses.  Using Eqs.~\eqref{eq:common_schw_response}--\eqref{eq:common_schw_noise} with $M\to M_I$, and assuming independent internal noises,
\begin{equation}
 \avg{\xi_I^{ij}(t)\xi_J^{kl}(t')}_{\rm st}
 =\delta_{IJ}\mathcal P_{\rm STF}^{ij,kl}N_I^{(0)}\delta(t-t').
 \label{eq:C_Schw_noise}
\end{equation}
For a stationary non-KMS state the same orbital transfer calculation goes through with the independently matched $\NN_I(\omega)$ in place of this white equilibrium coefficient.

For body $I$ in the Newtonian field of $J$ we choose the sign convention
\begin{equation}
\begin{aligned}
 \mathcal E^{(J)}_{ij}&=\frac{GM_J}{r^3}(3n_in_j-\delta_{ij}),\\
 K^{(J)}_{k|ij}\equiv\partial_k\mathcal E^{(J)}_{ij}
 &=\frac{3GM_J}{r^4}
 (\delta_{ki}n_j+\delta_{kj}n_i+\delta_{ij}n_k-5n_kn_in_j).
\end{aligned}
 \label{eq:C_tidal_vertex}
\end{equation}
Changing the overall sign convention for $\mathcal E_{ij}$ reverses the stochastic force but leaves all covariances below unchanged.  Thus $F^{(I)}_{\xi,k}=-K^{(J)}_{k|ij}\xi_I^{ij}$ and
$K_{k|ij}\mathcal P_{\rm STF}^{ij,lm}K_{l|lm}=18G^2M_J^2(\delta_{kl}+2n_kn_l)/r^8$.  The direct covariance is therefore
\begin{equation}
 \avg{F^{(I)}_{\xi,k}(t)F^{(I)}_{\xi,l}(t')}_{\rm st}
 =\frac{8}{5\pi}\frac{G^6M_I^5M_J^2}{r^8}
 (\delta_{kl}+2n_kn_l)\delta(t-t').
 \label{eq:C_direct_force_Schw}
\end{equation}

At this order, integrating out the potential metric at tree level gives the complete finite inter-body action that is linear in the stochastic electric quadrupoles,
\begin{equation}
 S_{\xi,\rm pot}^{\rm LO}=-\int\dd t\,
 \left[\xi_1^{ij}\mathcal E^{(2)}_{ij}+\xi_2^{ij}\mathcal E^{(1)}_{ij}\right].
 \label{eq:C_LO_stochastic_action}
\end{equation}
No second finite inter-body operator has the same field content and PN scaling in this truncation.  Varying Eq.~\eqref{eq:C_LO_stochastic_action} reproduces the direct term $B_z\xi_I$ together with the metric-mediated force obtained by sourcing the potential metric with $B_h\xi_I$, propagating it with $G_h^R$, and coupling it to the companion through $L_{zh}$; coincident propagation belongs to self-field renormalization.  This gives
\begin{equation}
 V_\xi(\bm r)=-\frac{G}{r^3}(3n_in_j-\delta_{ij})
 (M_2\xi_1^{ij}+M_1\xi_2^{ij}),\qquad
 \bm F_1^\xi=-\nabla_{\bm r}V_\xi=-\bm F_2^\xi .
 \label{eq:C_stochastic_binary_potential}
\end{equation}
For $M_{\rm tot}\equiv M_1+M_2$, $\mu=M_1M_2/M_{\rm tot}$ and independent horizon noises, the resulting full leading-PN acceleration-source covariance is
\begin{equation}
 \avg{a_{\xi,k}(t)a_{\xi,l}(t')}_{\rm st}^{\,\rm full}
 =C_0^{\rm full}(\delta_{kl}+2n_kn_l)\delta(t-t'),
 \quad
 C_0^{\rm full}=\frac{8}{5\pi}\frac{G^6M_{\rm tot}^2(M_1^3+M_2^3)}{r^8}.
 \label{eq:C_C0_full}
\end{equation}
By contrast, the fixed-metric reduction sets $h_\xi=0$ and keeps only $B_z\xi$, giving
\begin{equation}
 C_0^{\rm fix}=\frac{8}{5\pi}\frac{G^6M_1^2M_2^2M_{\rm tot}}{r^8},\qquad
 \frac{C_0^{\rm full}}{C_0^{\rm fix}}
 =\frac{(1+q)(1+q^3)}{q^2},\qquad q\equiv\frac{M_2}{M_1}.
 \label{eq:C_full_fixed_ratio}
\end{equation}
These formulas compare the same relative-acceleration source covariance at fixed background position.  They do not by themselves give the ratio of the final orbital covariance $C_{zz}$, which also depends on $G_{z,\rm eff}^R$.  For equal masses the acceleration variance is four times, and its root-mean-square amplitude twice, the fixed-metric result; for $q\ll1$ the variance ratio scales as $q^{-2}$.  The enhancement is a classical leading-PN coupled-feedback effect and does not alter the quantum scaling of the source.  In the fixed-metric reduction, action--reaction is not manifest because the metric-mediated force is omitted; in the coupled theory it follows automatically from the common translationally invariant action.

Restoring $\hbar$, Eq.~\eqref{eq:main_feedback_planck} shows that the variance is $\ord(\ell_{\rm P}^2)$ and the root-mean-square fluctuation is $\ord(\ell_{\rm P})$; the full leading-PN metric feedback changes only the classical coefficient.  For a stationary acceleration averaged over a time $\mathcal T$ longer than its correlation time but shorter than radiation reaction,
\begin{equation}
 \epsilon_{a,i}(\mathcal T)
 \equiv\frac{\sqrt{\avg{\bar a_i^2}_{\rm st}}}{r_{\rm tot}/r^2}
 =\sqrt{\frac{8\lambda_i}{5\pi}}\,
 \frac{\ell_{\rm P}\sqrt{r_1^3+r_2^3}}{r^2\sqrt{\mathcal T}},
 \qquad \lambda_r=3,\quad\lambda_\perp=1,
 \label{eq:C_acc_relative_error}
\end{equation}
so ordinary time averaging strengthens the suppression.

An integrated observable behaves differently.  For a conservative circular orbit, $\Omega^2=r_{\rm tot}/r^3$, the white acceleration noise gives, for $\Omega^{-1}\ll t\ll t_{\rm rr}$,
\begin{equation}
 \avg{\delta\phi^2(t)}_{\rm st}\simeq
 \frac{24}{5\pi}\ell_{\rm P}^2\frac{r_{\rm tot}^2(r_1^3+r_2^3)}{r^{10}}t^3,
 \qquad
 \epsilon_\phi(t)\equiv\frac{\sqrt{\avg{\delta\phi^2}_{\rm st}}}{\Omega t}
 \simeq\sqrt{\frac{24}{5\pi}}\,
 \frac{\ell_{\rm P}\sqrt{r_{\rm tot}(r_1^3+r_2^3)}}{r^{7/2}}\sqrt t .
 \label{eq:C_phase_relative_error}
\end{equation}
The secular $t^{1/2}$ growth makes the formal $t\to\infty$ limit nonuniform, with a crossover
$t_{\rm P}\sim(5\pi/24)r^7/[\ell_{\rm P}^2r_{\rm tot}(r_1^3+r_2^3)]$.  For a macroscopic binary this is parametrically beyond the conservative inspiral window, so the stochastic phase remains Planck suppressed over the regime in which the calculation is controlled.

\subsection{Scattering mass-shift variance: comparison with on-shell amplitude results}
\label{app:scattering_mass_variance}

The bound-orbit calculation above and the scattering mass shift probe the same local tidal vertex and black-hole noise kernel, but along different trajectories.  A coefficient-level comparison with Ref.~\cite{Aoki:2025quantum} therefore requires evaluating the EFT on the same scattering geometry rather than reusing the circular-orbit covariance.  At leading order in $G$ and in the nonrelativistic limit, take the undeflected encounter
\begin{equation}
 \bm r(t)=(b,vt,0),\qquad R(t)=\sqrt{b^2+v^2t^2},\qquad
 \mathcal E^{(2)}_{ij}(t)=\frac{GM_2}{R^3}(3n_in_j-\delta_{ij}),
 \label{eq:D_scattering_trajectory}
\end{equation}
with $\bm n=\bm r/R$.  The leading finite inter-body vertex in Eq.~\eqref{eq:C_LO_stochastic_action} gives the stochastic mass transfer to body 1,
\begin{equation}
 \delta\dot M_{1,\xi}(t)=\xi_1^{ij}(t)\dot{\mathcal E}^{(2)}_{ij}(t),\qquad
 \delta\Delta M_1=\int_{-\infty}^{\infty}\dd t\,
 \xi_1^{ij}(t)\dot{\mathcal E}^{(2)}_{ij}(t).
 \label{eq:D_scattering_mass_shift}
\end{equation}
Thus the integrated mass-shift variance, unlike the instantaneous acceleration covariance, contains the full encounter history,
\begin{align}
 \avg{(\delta\Delta M_1)^2}_{\rm st}
 &=\int\dd t\,\dd t'\,
 \dot{\mathcal E}^{(2)}_{ij}(t)
 N_1^{ij,kl}(t-t')
 \dot{\mathcal E}^{(2)}_{kl}(t') \notag\\
 &=\int\frac{\dd\omega}{2\pi}\,
 \widetilde{\dot{\mathcal E}}{}^{(2)}_{ij}(-\omega)
 N_1^{ij,kl}(\omega)
 \widetilde{\dot{\mathcal E}}{}^{(2)}_{kl}(\omega).
 \label{eq:D_scattering_mass_variance_general}
\end{align}
Here we keep $\hbar$ explicit.  For the leading soft Unruh contribution relevant to Ref.~\cite{Aoki:2025quantum}, the symmetric electric-quadrupole kernel is local,
\begin{equation}
 N_{1,U}^{ij,kl}(t-t')=
 \frac{4\hbar}{45\pi}G^4M_1^5\,
 \mathcal P_{\rm STF}^{ij,kl}\delta(t-t').
 \label{eq:D_Unruh_white_kernel}
\end{equation}
This is the same low-frequency local coefficient as Eq.~\eqref{eq:common_schw_noise}; no global KMS assumption for the Unruh state is needed for the comparison.  Along Eq.~\eqref{eq:D_scattering_trajectory},
\begin{equation}
 \dot{\mathcal E}^{(2)}_{ij}\dot{\mathcal E}^{(2)}_{ij}
 =18G^2M_2^2v^2\frac{b^2+3v^2t^2}{(b^2+v^2t^2)^5},\qquad
 \int_{-\infty}^{\infty}\dd t\,
 \dot{\mathcal E}^{(2)}_{ij}\dot{\mathcal E}^{(2)}_{ij}
 =\frac{225\pi}{32}\frac{G^2M_2^2v}{b^7}.
 \label{eq:D_Unruh_tidal_integral}
\end{equation}
Since $\mathcal E_{ij}$ is STF, Eqs.~\eqref{eq:D_scattering_mass_variance_general}--\eqref{eq:D_Unruh_tidal_integral} give
\begin{equation}
 \avg{(\delta\Delta M_1)^2}_{U}
 =\frac{5}{8}\frac{\hbar G^6M_1^5M_2^2}{b^7}\,v+\ord(v^3),
 \label{eq:D_Unruh_mass_variance}
\end{equation}
which is the nonrelativistic limit of Eq.~(4.45) in Ref.~\cite{Aoki:2025quantum}.

For the Boulware state, the Ohmic absorptive response implies the zero-temperature symmetric noise
\begin{equation}
 N_{1,B}^{ij,kl}(\omega)=\hbar\gamma_{\EE2,1}|\omega|\,
 \mathcal P_{\rm STF}^{ij,kl},\qquad
 \gamma_{\EE2,1}=\frac{16}{45}G^5M_1^6.
 \label{eq:D_Boulware_Ohmic_kernel}
\end{equation}
For the same straight-line encounter the remaining Fourier integral is
\begin{equation}
 \int\frac{\dd\omega}{2\pi}|\omega|\,
 \widetilde{\dot{\mathcal E}}{}^{(2)}_{ij}(-\omega)
 \widetilde{\dot{\mathcal E}}{}^{(2)}_{ij}(\omega)
 =\frac{192}{\pi}\frac{G^2M_2^2v^2}{b^8},
 \label{eq:D_Boulware_tidal_integral}
\end{equation}
so that
\begin{equation}
 \avg{(\delta\Delta M_1)^2}_{B}
 =\frac{1024}{15\pi}
 \frac{\hbar G^7M_1^6M_2^2}{b^8}\,v^2+\ord(v^4),
 \label{eq:D_Boulware_mass_variance}
\end{equation}
in agreement with the nonrelativistic limit of Eq.~(4.44) in Ref.~\cite{Aoki:2025quantum}.  The agreement in Eqs.~\eqref{eq:D_Unruh_mass_variance} and \eqref{eq:D_Boulware_mass_variance} therefore checks the local noise normalization after both descriptions are evaluated on the same scattering trajectory; it is not a geometry-independent equality between the bound and unbound observables.

\section{Superposed tidal source in DSW setup}
\label{app:decoherence}

We derive here the electric tidal $a$-value used in Section~\ref{subsec:decoherence_general}, keeping the numerical coefficients explicit.  Let
$\rho_a(\bm x)=\rho_1(\bm x)-\rho_2(\bm x)$ be the branch difference of the complete supported laboratory.  Equal total mass and fixed center of mass imply
\begin{equation}
  \int\dd^3x\,\rho_a=0,
  \qquad
  \int\dd^3x\,x^i\rho_a=0,
  \label{eq:D_mono_dipole}
\end{equation}
so the first nonzero far-zone moment is generically the STF mass quadrupole
\begin{equation}
  I_{a,ij}
  \equiv
  \int \dd^3x\,\rho_a(\bm x)\,x_{\langle i}x_{j\rangle}
  =\mathcal O(md^2).
  \label{eq:D_effective_quadrupole}
\end{equation}
At a black-hole position $\bm X=D\bm n$ with $D\gg d$, the quadrupolar potential difference is
\begin{equation}
  \Phi_a(\bm X)
  =-\frac{3G}{2D^3}\,I_{a,ij}n^in^j
  +\mathcal O(D^{-4}).
  \label{eq:D_quad_potential}
\end{equation}
In the weak-field vacuum region, $\EE_{a,ij}=\partial_{\langle i}\partial_{j\rangle}\Phi_a$.  Defining $I_{a,in}\equiv I_{a,ik}n^k$ and $I_{a,nn}\equiv I_{a,ij}n^in^j$, one obtains
\begin{align}
  \EE_{a,ij}
  =-\frac{3G}{2D^5}\Big[
  2I_{a,ij}
  -20 n_{(i}I_{a,j)n}
  -5\delta_{ij}I_{a,nn}
  +35n_in_jI_{a,nn}
  \Big]
  +\mathcal O(D^{-6}),
  \label{eq:D_tidal_from_quad}
\end{align}
where symmetrization has weight one half.  Contracting the tensor gives
\begin{align}
  \EE_{a,ij}\EE_a^{ij}
  =\frac{9G^2}{4D^{10}}\Big[
  4I_{a,ij}I_a^{ij}
  +120I_{a,in}I_a^{in}
  +90(I_{a,nn})^2
  \Big]
  +\mathcal O(D^{-11}).
  \label{eq:D_tidal_path_exact}
\end{align}
Thus $I_{a,ij}=\mathcal O(md^2)$ implies $\EE_a^2=\mathcal O(G^2m^2d^4/D^{10})$, with the complete coefficient and angular dependence displayed explicitly above.

We can also verify the normalization of Eq.~\eqref{eq:DSW_scaling_units}.  In $G=\hbar=c=1$ units, Eq.~\eqref{eq:common_schw_noise} and the STF property of $\EE_{a,ij}$ give
\begin{equation}
  \Gamma_{\rm dec}^{\rm BH}(t_f)
  =\frac{2M^5t_f}{45\pi}\,\EE_{a,ij}\EE_a^{ij}.
  \label{eq:D_DSW_E_exact}
\end{equation}
Substituting Eq.~\eqref{eq:D_tidal_path_exact} therefore yields
\begin{equation}
  \Gamma_{\rm dec}^{\rm BH}(t_f)
  =\frac{M^5t_f}{10\pi D^{10}}
  \left[
  4I_{a,ij}I_a^{ij}
  +120I_{a,in}I_a^{in}
  +90(I_{a,nn})^2
  \right].
  \label{eq:D_DSW_I_exact}
\end{equation}
Writing $q_{ij}\equiv I_{a,ij}/(md^2)$ and restoring dimensions gives
\begin{equation}
  \Gamma_{\rm dec}^{\rm BH}(t_f)
  =C_{\rm geom}
  \frac{G^6M^5m^2d^4}{\hbar c^{10}D^{10}}\,t_f,
  \qquad
  C_{\rm geom}
  =\frac{4q_{ij}q^{ij}+120q_{in}q^{in}+90q_{nn}^2}{10\pi}.
  \label{eq:D_DSWresult}
\end{equation}
Thus Eq.~\eqref{eq:DSW_scaling_units} is the geometry-independent scaling obtained after suppressing the dimensionless factor $C_{\rm geom}$.

\bibliographystyle{JHEP}
\bibliography{references}

\end{document}